\documentstyle[12pt]{article}
\makeatletter
\newdimen\normalarrayskip              
\newdimen\minarrayskip                 
\normalarrayskip\baselineskip
\minarrayskip\jot
\newif\ifold             \oldtrue            \def\new{\oldfalse}
\def\arraymode{\ifold\relax\else\displaystyle\fi} 
\def\eqnumphantom{\phantom{(\theequation)}}     
\def\@arrayskip{\ifold\baselineskip\z@\lineskip\z@
     \else
     \baselineskip\minarrayskip\lineskip2\minarrayskip\fi}
\def\@arrayclassz{\ifcase \@lastchclass \@acolampacol \or
\@ampacol \or \or \or \@addamp \or
   \@acolampacol \or \@firstampfalse \@acol \fi
\edef\@preamble{\@preamble
  \ifcase \@chnum
     \hfil$\relax\arraymode\@sharp$\hfil
     \or $\relax\arraymode\@sharp$\hfil
     \or \hfil$\relax\arraymode\@sharp$\fi}}
\def\@array[#1]#2{\setbox\@arstrutbox=\hbox{\vrule
     height\arraystretch \ht\strutbox
     depth\arraystretch \dp\strutbox
     width\z@}\@mkpream{#2}\edef\@preamble{\halign \noexpand\@halignto
\bgroup \tabskip\z@ \@arstrut \@preamble \tabskip\z@ \cr}%
\let\@startpbox\@@startpbox \let\@endpbox\@@endpbox
  \if #1t\vtop \else \if#1b\vbox \else \vcenter \fi\fi
  \bgroup \let\par\relax
  \let\@sharp##\let\protect\relax
  \@arrayskip\@preamble}
\def\eqnarray{\stepcounter{equation}%
              \let\@currentlabel=\theequation
              \global\@eqnswtrue
              \global\@eqcnt\z@
              \tabskip\@centering
              \let\\=\@eqncr
              $$%
 \halign to \displaywidth\bgroup
    \eqnumphantom\@eqnsel\hskip\@centering
    $\displaystyle \tabskip\z@ {##}$%
    &\global\@eqcnt\@ne \hskip 2\arraycolsep
         $\displaystyle\arraymode{##}$\hfil
    &\global\@eqcnt\tw@ \hskip 2\arraycolsep
         $\displaystyle\tabskip\z@{##}$\hfil
         \tabskip\@centering
    &{##}\tabskip\z@\cr}
\makeatother

\def\beq{\begin{equation}}
\def\eeq{\end{equation}}
\def\bea{\begin{eqnarray}}
\def\eea{\end{eqnarray}}

\def\Bf#1{\mbox{\boldmath $#1$}}

\def\balpha{{\Bf\alpha}}
\def\bbeta{{\Bf\beta}}

\def\bmu{{\Bf\mu}}
\def\bphi{{\Bf\phi}}
\def\bN{{\Bf N}}

\def\bJ{{\Bf J}}

\def\bx{{\Bf x}}

\def\bsalpha{{\Bf\alpha}}
\def\bsbeta{{\Bf\beta}}

\def\bsphi{{\Bf\phi}}
\def\bsN{{\Bf N}}

\def\bsx{{\Bf x}}

\def\W{{\rm W}}

\def\nn{\nonumber}
\def\mm{matrix model }

\def\stackreb#1#2{\mathrel{\mathop{#2}\limits_{#1}}}
\begin{document}

\begin{titlepage}
\begin{center}
\begin{flushright}{NORDITA-93/21 \ \ FIAN/TD-05/93}\end{flushright}
\begin{flushright}{March 1993}\end{flushright}
\vspace{0.1in}{\Large\bf Integrable structures in matrix models\\
and physics of 2d-gravity
\footnote{based on lectures given at the Niels Bohr
Institute}
}\\[.4in]
{\large  A. Marshakov
\footnote{E-mail address: marshakov@nbivax.nbi.dk \ \ tdparticle@glas.apc.org}
}\\
\bigskip {\it NORDITA, Blegdamsvej 17, DK-2100,
Copenhagen {\O}, Denmark\\and\\Theory Department\\ P.N.Lebedev Physics
Institute \\ Leninsky prospect, 53, Moscow, 117 924, Russia
\footnote{permanent address}
}

\end{center}
\bigskip \bigskip

\begin{abstract}
A review of the appearence of integrable structures in
the matrix model desription of $2d$-gravity is presented. Most of the ideas
are
demonstrated at the technically simple but ideologically important
examples.
Matrix models are
considered as a sort of "effective" description of continuum $2d$ field
theory formulation. The main
physical role in such description is played by the Virasoro-$W$
conditions, which can be interpreted as certain unitarity or factorization
constraints. Both discrete and continuum (Generalized Kontsevich) models
are formulated as the solutions to those discrete (continuous)
Virasoro-$W$ constraints. Their integrability properties are proved,
using mostly the determinant technique highly related to the
representation in terms of free fields. The paper also contains some new
observations connected to formulation of more general than GKM solutions
and deeper understanding of their relation to $2d$ gravity.
\end{abstract}

\end{titlepage}

\newpage
\setcounter{footnote}0

\section{Introduction}

The most geometrical way to derive the $2d$ gravity or string theory partition
function (as well as the generating function for the correlators) is
given by the Polyakov path integral which for the partition function
reads

\beq\label{pol}
F(\lambda) = \sum_{p-genus} \lambda ^p F_p
$$
$$
F_p = \int _{\Sigma_p} Dg \exp \gamma \int R {\Delta}^{-1} R
\eeq

Matrix models historically appeared \cite{mamo} when one considers
the
discretization of (\ref{pol}) which in a sense turns to be exact at
least for the simplest cases of ''empty" string theories, when the
target space is low-dimensional (or in the limit of pure gravity does not
exist at all). A well-known example of such theories is given by minimal
$(p,q)$ models coupled to $2d$ gravity.

The difficulties with the continuum formulation (\ref{pol}) as usual
arise from the fact that it possesses additional complicated information
(like the ''Verma-module" structure of the underlying $2d$ CFT) which might
not be essential for description of final ''effective" theory. We have
exactly here the case of a gauge-theory, when after coupling to gravity
all conformal descendants are ''gauged away" and one might search for a
sort of ''frame" description, which luckily appears in the form of
matrix models.

Matrix models are usually defined by the integrals like

\beq\label{mamo}
Z_N = \int DM_{N \times N} \exp [- TrV(M)]
\eeq
which in the continuum limit, requiring in part $N \to \infty$

\beq\label{conlim}
\log Z \stackreb{N \to \infty}{\to} F
\eeq
can give the whole {\it nonperturbative} solution to (\ref{pol})
\cite{Ketal}.

However, below we will mostly advocate the point of view different from
the original definition (\ref{mamo}). Indeed, the main difference
between the continuum (\ref{pol}) and the \mm formulation is that the
first one requires some sort of the unitarity or factorization
relations to connect different terms in the sum over topologies in
(\ref{pol}) while in the \mm formulation (\ref{mamo}) these relations
usually appears {\it automatically}. Moreover, for known solutions they
usually appear in the well-known form of the Virasoro-$W$
\footnote{This should have an interpretation as a sort of world-sheet --
target-space duality between the Virasoro symmetry as gauge symmetry for
$2d$ gravity and Virasoro relations in the target-space (or better in
the space of coupling constants).}
constraints,
which in a sense may be considered as a {\it definition} of \mm.

Below, we will usually start the description of various \mm and \mm-like
theories as being particular solutions to these Virasoro-$W$ recursion
relations. It turns out that these relations lead to the {\it
integrability} property of matrix models, namely the solutions to these
constrains
turn to be $\tau$-functions of the hierarchies of well-known integrable
equations \cite{Douglas,FKN1,DVV91a}.

In terms of the partition function (\ref{pol}), (\ref{conlim})

\beq\label{tau}
F(T) = \log \tau (T)
\eeq
where $T \equiv \{T_k\}$ is the set of {\it times} or the coupling
constants in the theory of $2d$ gravity.

The formula (\ref{tau}) or the appearence of the integrability is
exactly what allows one to make more progress in studying ''frame"
formulation (\ref{mamo}) instead of the original one (\ref{pol}).

In what follows, we will first consider an example of {\it discrete} matrix
models ({\it
finite} $N$ in (\ref{mamo})) as being solution to the most simple
discrete Virasoro constraints and then pass to the continuum case. Both
kinds of solutions to the recursion relations are $\tau$-functions of
well-known
hierarchies and both particular solutions have representation in
{\it integral} form. We will
also stress the moment that the solutions to discrete constraints
correspond to the discretization of the world-sheet of string, while, as
it will be seen below the matrix solutions to the continuum relations
have rather interpretation of target-space theory or of an effective
{\it string field theory}.

\section{Discrete hermitean 1-matrix model as a solution
to the discrete Virasoro constraints}

In this section we are going to consider the first (and simplest) example,
demonstrating the above ideas, namely
the solution to the discrete Virasoro constraints \cite{MMM91a}:
$$
L_nZ[t] = 0,\ \  \   n \geq  -1
$$
\beq\label{dvir}
L_n \equiv \sum ^\infty _{k=0}kt_k\partial /\partial t_{k+n} +
\sum _{a+b=n}\partial ^2/\partial t_a\partial t_b
\eeq
with an additional requirement (concerning $t_0$-variable)
$$
\partial Z_{N}/\partial t_0 = -NZ_{N}
$$
(later on $N$ will be identified with the size of
matrices in the formulas like (\ref{mamo})).
The key idea how to solve the constraints (\ref{dvir}) appears after one
notices that the Virasoro generators (\ref{dvir})
actually have the well-known form of the Virasoro operators in
the theory of one free scalar field
\footnote{In this sense we have $c=1$ for (\ref{dvir}) though it is not
too sensible to speak of the central charge, having only the {\it half}
($n \geq -1$) of the Virasoro algebra, i.e.
$$
[L_n,L_m] = (n-m)L_{n+m}\ , \ n,m \geq -1
$$
always {\it without} the central element.}.
If we try to look for such solution in terms of
{\it holomorphic} components of the scalar field
\bea\label{freesc}
\phi (z) &=  \hat q + \hat p \log z  + \sum _{k\neq 0} {J_{-k}\over k}
z^{-k}\nn\\
\  [J_n,J_m] &= n\delta _{n+m,0},  \ \ \     [\hat q,\hat p] = 1
\eea
the procedure is as follows. First, define the vacuum states
\bea\label{bosvac}
J_k|0\rangle  &= 0, \ \ \  \langle N|J_{-k} = 0, \ \ \    k > 0\nn\\
\hat p|0\rangle  &= 0, \ \ \   \langle N|\hat p = N\langle N|
\eea
''Half" of the stress-tensor\footnote
{For the sake of brevity, we omit the sign of normal ordering in the
evident places, say, in the expression for $T$ and $W$ in terms of
free fields.} components
\beq\label{tensor}
T(z) = {1\over 2}[\partial \phi (z)]^2 = \sum    T_nz^{-n-2},\quad
T_n = {1\over 2}\sum _{k>0}J_{-k}J_{k+n} +
{1\over 2}\sum _{{a+b=n}\atop{a,b\geq 0}}J_aJ_b,
\eeq
obviously vanish the $SL(2)$-invariant vacuum
\beq\label{6a}
 T_n|0\rangle  = 0,  \ \ \    n \geq  -1
\eeq
Second, we define the Hamiltonian by
\bea\label{ham}
H(t) &= {1\over \sqrt{2}} \sum _{k>0}t_kJ_k =
\oint_{C_0}V(z)j(z)\nn\\
V(z) &= \sum _{k>0}t_kz^k, \ \  \   j(z) = {1\over \sqrt{2}}\partial \phi (z).
\eea
Now one can easily construct a ``conformal field theory" solution to
(\ref{dvir}) in two steps. The basic ''transformation"
\beq\label{trick}
L_n\langle N|e^{H(t)}\ldots = \langle N|e^{H(t)}T_n\ldots
\eeq
can be checked explicitly. As an immediate consequence, any correlator of the
form
\beq\label{dcor}
\langle N|e^{H(t)}G|0\rangle
\eeq
($N$  counts the number of zero modes, ''included" in  $G$ -- that is the
role of the {\it size} of matrix in (\ref{mamo})) gives a
solution to
(\ref{dvir})
provided by
\beq\label{com}
[T_n,G] = 0, \ \ \  n \geq  -1
\eeq
The conformal solution to (\ref{com})
(and therefore to (\ref{dvir})) immediately comes from the basic
properties of $2d$ conformal algebra. Indeed, any solution to
\beq\label{comm}
[T(z),G] = 0
\eeq
is a solution to (\ref{com}),
and it is well-known that the solution to (\ref{comm}) is (by definition
of the chiral algebra) a
function of {\it screening charges} in the free scalar field theory
given by
\beq\label{scr}
Q_\pm  = \oint J_\pm  = \oint
e^{\pm \sqrt{2}\phi }.
\eeq
With a selection rule on zero mode it gives
\beq\label{discrG}
G = \exp \ Q_+ \rightarrow  {1\over N!}Q^N_+
\eeq
(Of course, the general case might be  $G \sim  Q^{N+M}_+Q^M_-$ but the special
prescription for integration contours, proposed in \cite{MMM91a},
implies that the
dependence of $M$ can be irrelevant and one can just put  $M = 0.)$
In this case the solution
\beq\label{z2cor}
Z[t] \equiv Z_{2,N}[t] = \langle N|e^{H(t)}\exp Q_+|0\rangle
\eeq
after computation of the free theory correlator, analytic continuation of the
integration contour
gives the result
\bea\label{z2}
Z_{2,N} &= (N!)^{-1}\int   \prod ^N_{i=1}dz_i \exp  \left( - \sum
t_kz^k_i\right)  \Delta ^2_N(z) =\nn\\
&= (N!{\rm Vol}\ U(N))^{-1}\int   DM\ \exp \left( - \sum    t_kM^k\right)
\\
\Delta _N &= \prod ^N_{i<j}(z_i - z_j)\nn
\eea
in the form of {\it multiple} integral over the ''spectral parameter"
or, in this particular case, in the form of the integral over
Hermitian matrices of the type of eq.(\ref{mamo}).

This point of view actually could be considered as a constructive one.
Namely, instead of
considering a special direct multi-matrix generalization of (\ref{z2})
one
can use powerful tools of $2d$ conformal field theories, where it is well
known how to
generalize almost all the steps of above construction: first, instead of
looking for a solution to Virasoro constraints one can impose {\it extended
Virasoro} or  $W$-constraints on the partition function. In such case one would
get Hamiltonians in terms of {\it multi}-scalar field theory, and the second
step is generalized directly using {\it screening charges} for  $W$-algebras.
The general scheme of solving discrete $W$-constraints looks as
follows \cite{KMMMP}:

(i)  Consider Hamiltonian as a linear combination of the Cartan currents of a
level one Kac-Moody algebra  ${\cal G}$
\beq\label{hamil}
H(t^{(1)},\ldots,t^{({\rm rank}\ {\cal G})}) =
\sum _{\lambda ,k>0}t^{(\lambda )}_k\bmu _\lambda \bJ_k,
\eeq
where $\{\bmu_i\}$ are basis vectors in Cartan hyperplane, which
for $SL(p)$ case are chosen to satisfy
$$
\bmu_i\cdot \bmu_j=\delta_{ij}-{1\over p},  \ \ \ \sum_{j=1}^p
\bmu_j=0.
$$

(ii)  The action of differential operators  ${\cal W}^{(a)}_i$ with respect to
times  $\{t^{(\lambda )}_k\}$ can be now defined from the relation
\beq\label{wtrick}
W^{(a)}_i\langle \bN|e^{H(\{t\})}\ldots =
\langle \bN|e^{H(\{t\})}{\rm W}^{(a)}_i\ldots\ , \ \ \   a=2,\ldots,p;  \ \ \
i\geq 1-a,
\eeq
where
\bea\label{wmodes}
\W^{(a)}_i &= \oint z^{a+i-1}\W^{(a)}(z)\nn\\
\W^{(a)}(z) &= \sum  _\lambda  [\bmu _\lambda \partial \bphi (z)]^a + \ldots
\eea
are  spin-$a$ W-generators of  $\W_p$-algebra written in terms of
rank$\;{\cal G}$-component scalar fields \cite{FL88}.

(iii)  The conformal solution to the discrete $W$-constraints arises in the
form \cite{KMMMP}
\beq\label{zcmm}
Z^{\rm CMM}_{p,\bsN}[\{t\}] = \langle \bN|e^{H(\{t\})}G\{Q
^{(\alpha)} \}|0\rangle
\eeq
where  $G$  is again an exponential function of screenings of level one
Kac-Moody algebra (see \cite{GMMOS} and references therein)
\beq\label{screen}
Q^{(\alpha)}  = \oint J^{(\alpha)}  = \oint e^{\bsalpha \bsphi }
\eeq
$\{\balpha \}$ being roots of finite-dimensional simply laced
Lie algebra ${\cal G}$.
The correlator (\ref{zcmm}) is still a free-field correlator
and the computation gives it again in a multiple integral form
\bea\label{multi}
Z^{\rm CMM}_{p,\bsN}[\{t\}] &\sim  \int   \prod  _\alpha
\left[ \prod ^{N_\alpha }_{i=1}dz^{(\alpha )}_i \exp \left( -
\sum _{\lambda ,k>0}t^{(\lambda )}_k(\bmu _\lambda \balpha )(z^{(\alpha )}_i)^k
\right) \right] \times \nn\\
&\times \prod _{(\alpha ,\beta )}\prod ^{N_\alpha }_{i=1}
\prod ^{N_\beta }_{j=1}(z^{(\alpha )}_i- z^{(\beta )}_j)^{\bsalpha \bsbeta }
\eea
The only difference with the one-matrix case (\ref{z2})
 is that the
expressions (\ref{multi}) have rather complicated representation in terms of
multi-matrix integrals, the following objects will necessarily appear
\bea\label{matten}
\prod ^{N_\alpha }_{i=1} \prod ^{N_\beta }_{j=1}(z^{(\alpha )}_i-
z^{(\beta )}_j)^{\bsalpha \bbeta } = \left[ \det \{M^{(\alpha )}\otimes I -
I\otimes M^{(\beta )}\}\right] ^{\bsalpha \bsbeta },
\eea
However, this is still a model with a chain of matrices
and with closest neibour interactions only (in the case of $SL(p)$).

Actually, it can be shown that CMM, defined by (\ref{zcmm}) as a
solution to the $W$-constraints has a very rich integrable structure
and possesses a natural continuum limit \cite{KMMMP,MP}. To pay for these
advantages one
should accept a slightly less elegant matrix integral with the entries like
 (\ref{matten}).

The first non-trivial example is the $p=3$ solution to
$W_3$-algebra: an alternative to the conventional 2-matrix model.
In this case
one has six screening charges  $Q^{(\pm \alpha _i)}$ $(i = 1,2,3)$
which commute with
\beq\label{w2}
\W^{(2)}(z) = T(z) = {1\over 2}[\partial \bphi (z)]^2
\eeq
and
\beq\label{w3}
\W^{(3)}(z) = \sum ^3_{\lambda =1}(\bmu _\lambda \partial \bphi (z))^3,
\eeq
where  $\bmu _\lambda $ are vectors of one of the fundamental representations
({\bf 3} or $\bar {\bf 3})$ of  $SL(3)$.

The particular form of integral representation (\ref{multi})
depends on particular screening insertions to the correlator (\ref{zcmm}).
We will concentrate on the solutions which have
no denominators. One of the reasons of
such choice is that these solutions possess the most simple integrable
structure,
though the other ones can still be analyzed in the same manner.

The simplest solutions which have no denominators correspond to specific
correlators
\beq
Z^{\rm CMM}_{p,\bsN}[\{t\}] = \langle \bN|e^{H(\{t\})}\prod_i
\exp Q_{\alpha _i}|0\rangle
\eeq
when we take  $\alpha _i$ to be ``neighbour" (not simple!) roots:
$(\balpha _i\balpha _j) =
1$. In the case of  $SL(3)$  this corresponds, say, to  insertions of only
  $Q_{\alpha _1}$ and  $Q_{\alpha _2}$ (again, we use the notations of
\cite{GMMOS}) and gives
\bea\label{zsl3}
&Z^{\rm CMM}_{3;M,N}[t,\bar t] \equiv  Z_{M,N}[t,\bar t] = {1\over N!M!}
\langle N,M|e^{H(t,\bar t)}(Q^{(\alpha _1)})^N(Q^{\alpha _2})^M|0\rangle  =
\nn\\
&= {1\over N!M!}\int   \prod    dx_idy_i \exp \left( - \sum    [V(x_i) +
\bar V(y_i)]\right)  \Delta ^2_N(x) \Delta ^2_M(y) \prod _{i,j}(x_i - y_j)
\eea

Possible generalizations of the above scheme can include
the ''supersymmetric matrix models" \cite{AG91},
 where the authors looked for a solution to the system of equations
 $L_nZ = 0$  and  $G_mZ = 0$,  with generators $\{L_n,G_m\}$  forming  the
$N = 1$  superconformal algebra, which
in our language is nothing but a trivial generalization of the
one-field case which has to be substituted by a scalar superfield.
Then the insertion of screenings of  $N = 1$  superconformal algebra
immediately leads to the result of \cite{AG91}.
{}From this point of view the real problem with supersymmetric generalization
can arise only in the  $N = 2$ case because of the lack of appropriate
screening operators.

\subsection{Determinant representation and integrability of the
solutions to Virasoro- $W$ constraints}

In the simplest case of the Hermitean one-matrix model the $N\times N$ matrix
integral (\ref{z2})
is taken by orthogonal polynomials for the arbitrary potential  $V(m) = \sum
t_km^k$

\beq\label{sp}
\langle i|j \rangle \equiv \langle P_i,P_j \rangle = \int
P_i(m)P_j(m)e^{-V(m)}dm =
\delta _{ij}e^{\varphi _i(t)}
\eeq
and equals

\beq\label{tautc}
Z_N = \prod ^{N-1}_{i=0}e^{\varphi _i(t)}
\eeq
It follows from (\ref{sp})
and the definition of orthogonal polynomials

\beq\label{defpol}
\new
\begin{array}{c}
P_i(m) = \sum _{j\leq i}a_{ij}m^j \\
a_{ii} = 1
\end{array}
\eeq
that

\beq\label{gauss}
diag(e^{\varphi _i(t)}) = A H A^{T} ,
\eeq
where  $A = \|a_{ij}\|$,  $A^T$ -- transponed matrix, and  $H$
is so called matrix of moments

\beq\label{moments}
H_{ij}= \int   m^{i+j}e^{-V(m)}dm
\eeq
Thus,

\beq\label{detmom}
Z _N[t] = \det [\hbox{diag}(e^{\varphi _i(t)})] =
\det \ AHA^{T} = \det \ H = \tau _N[t]
\eeq

Without going into all the details, which could be found in
\cite{GMMMO,KMMOZ} we will only point out that (\ref{gauss}) is a kind
of a Riemann-Hilbert problem and the determinant formula (\ref{detmom})
is {\it one of the basic definitions} of $\tau$-function of Toda theory.

The relation (\ref{tautc}) and (\ref{detmom}) is actually based only on the
fact that in the
theory of Toda-chain there exists a relation between the potentials of
the Toda-chain equations and the $\tau$-function having the form of a
{\it difference} operator:

\beq\label{diftc}
e^{\varphi _i (t)} = {\tau _{i+1} (t) \over \tau _i (t)}
\eeq

Now to check that we have really got Toda chain hierarchy, let us check
the first equation - flow in the direction $\partial \over \partial
t_1$. This might be done after we introduce the {\it Lax} operator for
the Toda chain (see \cite{GMMMO} for details), which in the basis of the
orthogonal polynomials (\ref{sp}) acts by:

\beq\label{lax}
mP_i(m) = P_{i+1}(m) - p_i(t)P_i(m) + R_i(t)P_{i-1}(m)
\eeq
$i.e.$ is determined by a {\it trilinear} matrix, what follows from a sort of
''unitarity
condition" or just the properties of the basis (\ref{sp}).
Now, from (\ref{sp}), (\ref{lax}) one can easily establish the relations
among the ''potentials" $\{\varphi _i(t) \}$ and the matrix elements
$\{ R_i(t),p_i(t) \}$, first:

\beq
\langle i|m|i-1 \rangle = e^{\varphi _i (t)} = R_i(t) e^{\varphi
_{i-1}(t)}
\eeq
gives

\beq\label{rfi}
R_i(t) = e^{\varphi _i(t) - \varphi _{i-1}}
\eeq
Now, differentiating (\ref{sp}) for $i=j$ one gets

\beq\label{p_i}
{\partial \over \partial t_1} \langle i|j \rangle = e^{\varphi _i(t)}
{\partial \varphi _i \over \partial t_1} = \int dm e^{-\sum t_k m^k}
(-mP^2_i + 2P_i {\partial P_i \over \partial t_1}) = p_i(t)e^{\varphi
_i(t)}
\eeq
where the second term in brackets dissappear again from the orthogonality
condition and property (\ref{defpol}).
{}From (\ref{p_i}) it follows that

\beq\label{moment}
p_i(t) = {\partial \varphi _i (t) \over \partial t_1}
\eeq
is just the {\it momentum} for the $\varphi _i (t)$-coordinate.

Differentiating (\ref{sp}) with $i > j$, we obtain

\beq
0 = - \int dm e^{-\sum t_k m^k} \left( mP_i P_j + P_j {\partial P_i \over
\partial t_1} \right)
\eeq
comparing which with (\ref{lax}) and using (\ref{sp}) one gets

\beq\label{piri}
{\partial P_i \over \partial t_1} = R_i P_{i-1}
\eeq
Now we are ready to differentiate (\ref{lax})

\beq\label{dlax}
m{\partial P_i \over \partial t_1} = {\partial P_{i+1} \over \partial
t_1} - {\partial p_i \over \partial t_1} P_i + ...
\eeq
Multiplying (\ref{dlax}) by $P_i$ integrating and using (\ref{sp}) and
(\ref{piri}) one finally gets

\beq
{\partial p_i \over \partial t_1} = R_{i+1} - R_i
\eeq
or using (\ref{moment}), (\ref{rfi})

\beq\label{1todch}
{\partial ^2 \varphi _i \over \partial t_1^2} = e^{\varphi
_{i+1}-\varphi _i} - e^{\varphi _i - \varphi _{i-1}}
\eeq
which is nothing but the first Toda-chain equation. In terms of the
$\tau$-function (\ref{1todch}) can be rewritten in the {\it Hirota bilinear}
form:

\beq\label{hirota}
\tau _N(t) {\partial ^2\over \partial t^2_1}\tau _N(t) -
\left( {\partial \tau _N(t)\over \partial t_1}\right) ^2 =
\tau _{N+1}(t)\tau _{N-1}(t)
\eeq

The $\tau$- function of $p=2$ case (\ref{detmom})
 can be also written in the form
\beq\label{todachain}
Z_{2,N}(t) = \det _{N\times N}[\partial ^{i+j-2}C(t)] = \tau _N(t)
\eeq
with
\beq\label{todachain2}
\partial _{t_n}C(t) = \partial ^n_{t_1}C(t)
\eeq
Eq.(\ref{todachain2})
means that  $C(t)$  just has an {\it integral} representation
\beq\label{tcentry}
C(t) = \int   d\mu (z) \exp  \sum    t_kz^k,
\eeq
where $d\mu $ is some measure; the Virasoro constraints
(\ref{dvir}) fix
 the concrete measure ($d\mu = dz$) and the contour of integration in
(\ref{tcentry}).
The determinant form (\ref{todachain}) is an explicit
manifestation of the fact that the partition function does satisfy the Hirota
bilinear relations, the simplest one of which in this particular case takes the
form (\ref{hirota}).

Now one can generalize (\ref{todachain}) and (\ref{todachain2})
\cite{KMMMP}.
In the case of the $p=3$ model (\ref{zcmm}) we have to introduce two functions
instead of (\ref{tcentry}):
\beq\label{3entry}
C(t) = \int   dz\ \exp  [-V(z)],  \ \ \    \bar C(\bar t) = \int   dz\ \exp
[-\bar V(z)]
\eeq
where
$$
V(z) = \sum _{k>0}t_kz^k, \ \ \     \bar V(z) = \sum _{k>0}\bar t_kz^k
$$
and
\beq\label{3toda}
\partial _{t_n}C(t) = \partial ^n_{t_1}C(t), \ \ \
\partial _{\bar t_n}
\bar C(\bar t) = \partial ^n_{\bar t_1}\bar C(\bar t)
\eeq
Again, we can get the determinant representation, now having the form
$(\partial  \equiv
\partial /\partial t_1$,  $\bar \partial  \equiv  \partial /\partial \bar t_1)$
$$
Z_{N,M}(t,\bar t) =
$$
$$
= \det \left[{
\begin{array}{cccccccc}
C&\partial C&\ldots&\partial^{N-1}C&
\bar C&\bar\partial \bar C&\ldots&
\bar\partial^{M-1}\bar C\\
\partial C&\partial^{2}C&\ldots&\partial^{N}C&
\bar\partial \bar C&\bar\partial^{2}\bar C&\ldots&
\bar\partial^{M}\bar C\\
&\phantom{a}&&&&&&\\
\partial^{N+M-1}C&\partial^{N+M}C&\ldots&\partial^{2N+M-2}C&
\bar\partial^{N+M-1}\bar C&\bar\partial^{N+M}\bar C&\ldots&
\bar\partial^{2N+M-2}\bar C
\end{array}}\right]=
$$
\beq\label{3det}
\equiv  \tau _{N,M}(t,\bar t)
\eeq
which is exactly the double-Wronskian representation of a
$\tau $-function \cite{Hir}.

{}From representation (\ref{3det}) it is easy to derive the analogs of
the Hirota relation (\ref{hirota})
\beq\label{3hir}
{\partial ^2\over \partial t_1\partial \bar t_1}\log \tau _{N,M}(t,\bar t) =
{\tau _{N+1,M-1}(t,\bar t)
\tau _{N-1,M+1}(t,\bar t)\over \tau ^2_{N,M}(t,\bar t)}
\eeq

''Higher-times" Hirota relations have more complicated form.

\subsection{Fermionic representation}

Now we shall proceed to the representation of the solutions to the
Virasoro-$W$ constraints in terms of the
fermionic correlation functions
\footnote{Indeed, the two types of technique we are using are
practically equivalent: the symmetry (better {\it anti}symmetry) of
determinants under permutations is what reflects the anticommuting
nature of the fermions} developed for generic integrable systems in
\cite{DJKM}.

Indeed, the $\tau $-function of 2-component KP
hierarchy defined by the fermionic correlator
\beq\label{3.2.1}
\tau ^{(2)}_{N,M}(x,y) = \langle N,M|e^{H(x,y)}G|N+M,0\rangle
\eeq
with
\beq\label{3.2.2}
H(x,y) = \sum _{k>0}(x_kJ^{(1)}_k + y_kJ^{(2)}_k)
\eeq
\beq\label{3.2.3}
J^{(i)}(z) = \sum    J^{(i)}_kz^{-k-1} = {:}\psi
^{(i)}(z) \psi ^{(i)\ast }(z){:}
\eeq
\beq\label{3.2.4}
\psi ^{(i)}(z) \psi ^{(j)\ast }(z') = {\delta _{ij}\over z - z'} + \ldots\ \ .
\eeq
is equivalent to (\ref{z2cor}) for certain
$G$ when (\ref{3.2.1}) depends only on the differences  $x_k-y_k$. To
prove
this we have to make use of the free-fermion representation of  $SL(2)_{k=1}$
Kac-Moody algebra:
\bea\label{3.2.5}
&J_0 = {1\over 2}(\psi ^{(1)}\psi ^{(1)\ast } - \psi ^{(2)}\psi ^{(2)\ast }) =
{1\over 2}(J^{(1)} - J^{(2)})\nn\\
&J_+ = \psi ^{(2)}\psi ^{(1)\ast } \quad J_- = \psi ^{(1)}\psi ^{(2)\ast }
\eea
Now let us take $G$ to be the following exponent of a quadratic form
\beq\label{3.2.6}
G \equiv  {:}\exp  \left( \int   \psi ^{(2)}\psi ^{(1)\ast }\right){:}
\eeq
The only term which contributes into the correlator (\ref{3.2.1})
due to the charge
conservation rule is:
\beq\label{3.2.7}
G_{N,M} \equiv  G_{N,-N}\delta_{M,-N} = {1\over N!}\ {:}\left( \int
\psi ^{(2)}\psi ^{(1)\ast }\right) ^N{:}~\delta_{M,-N}
\eeq
Now one can bosonize the fermions
\bea\label{3.2.8}
\psi ^{(i)\ast } = e^{\phi _i},  \ \ \  \psi ^{(i)} = e^{- \phi _i}
\nn\\
J^{(1)} = \partial \phi _1, \ \ \     J^{(2)} = \partial \phi _2
\eea
and compute the correlator
$$
\tau ^{(2)}_N(x,y) \equiv  \tau ^{(2)}_{N,-N}(x,y) = {1\over N!}
\langle N,-N|\exp \left( \sum _{k>0}(x_kJ^{(1)}_k +
y_kJ^{(2)}_k)\right) \left( \int
{:}\psi ^{(2)}\psi ^{(1)\ast }{:}\right) ^N|0\rangle  =
$$
$$
= {1\over N!} \langle N,-N|\exp \left(
\oint
[X(z)J^{(1)}(z) + Y(z)J^{(2)}(z)]\right) \left( \int
:\exp (\phi _1-\phi _2):\right) ^N|0\rangle
$$
Introducing the linear combinations  $\sqrt{2}\phi  = \phi _1 - \phi _2$,
$\sqrt{2}\tilde \phi  = \phi _1 + \phi _2$ we finally get
\bea\label{3.2.9}
&\tau ^{(2)}_N(x,y) = {1\over N!} \langle \exp \left( {1\over \sqrt{2}}
\oint[X(z)+Y(z)]\partial \tilde \phi (z)\right) \rangle  \times\nn\\
&\times  \langle N|\exp \left( {1\over \sqrt{2}}
\oint [X(z)-Y(z)]\partial \phi (z)\right) \left( \int
:\exp \sqrt{2}\phi :\right) ^N|0\rangle  = \tau ^{(2)}_N(x-y)
\eea
since the first correlator is in fact independent of $x$ and $y$. Thus, we
proved that the $\tau $-function (\ref{3.2.1})
indeed depends only on the {\it difference} of two sets
of times  $\{x_k-y_k\}$, and coincides with (\ref{z2cor}).

The above simple example already contains all the basic features of at least
all the  $A_p$ cases. Indeed, the reduction (\ref{3.2.9})
is nothing but  $SL(2)$-reduction
of a generic  $GL(2)$  situation. In other words, the diagonal  $U(1)$
$GL(2)$-current  $\tilde J = {1\over 2}(J^{(1)} + J^{(2)}) =
{1\over \sqrt{2}}\partial \tilde \phi $  decouples. This is an invariant
statement which can be easily generalized to higher  $p$  cases.

In the case of  $SL(p)$  we have to deal with the $p$-component hierarchy and
instead of (\ref{3.2.1}) for generic $\tau $-function one has
\bea\label{3.2.10}
\tau ^{(p)}_N(x) &= \langle \bN|e^{H(x)}G|0\rangle\\
\bN &= \{N_1,\ldots,N_p\}, \ \ \     x = \{x^{(1)},\ldots,x^{(p)}\}\nn
\eea
and now we have  $p$  sets of fermions  $\{\psi ^{(i)\ast },\psi ^{(i)}\}$
$i=1,...,p$. The Hamiltonian is given by the Cartan currents of $GL(p)$
\bea\label{3.2.11}
H(t) &= \sum ^p_{i=1} \sum _{k>0}x^{(i)}_kJ^{(i)}_k\\
J^{(i)}(z) &= \psi ^{(i)}\psi ^{(i)\ast }(z)\nn
\eea
and the element of the Grassmannian in the particular case of CMM is given by
an exponents of the other currents
\beq\label{3.2.12}
J^{(ij)} = \psi ^{(i)}\psi ^{(j)}, \ \ \  \tilde J^{(ij)} =
\psi ^{(i)}\psi ^{(j)\ast },  \ \ \  J^{(ij)\ast } =
\psi ^{(i)\ast }\psi ^{(j)\ast },  \ \ \    i \neq  j
\eeq
i.e.
\bea\label{3.2.13}
G &\equiv  \prod     \exp (Q^{(ij)})\exp (\tilde Q^{(ij)})\exp (Q^{(ij)\ast })
\\
Q^{(ij)} &=  \oint
J^{(ij)},  \  \ \   \tilde Q^{(ij)} =\oint
\tilde J^{(ij)}, \ \ \   Q^{(ij)\ast } = \oint
J^{(ij)\ast }, \ \ \       i \neq  j \nn
\eea
Since  (\ref{3.2.12}) are the $GL(p)_1$ Kac-Moody currents, (\ref{3.2.13}) play
the role of screening operators in the theory.
It deserves mentioning that they are exactly  the $SL(p)$ (not  $GL(p))$
-screenings and thus the $\tau $-function (\ref{3.2.10}) does not depend on
$\{\sum ^p_{i=1}x^{(i)}_k\}$.

In the case of  $SL(3)$  this looks as follows. The screenings are
\beq\label{3.2.14}
Q^{(\bsalpha )} =\oint
J^{(\bsalpha )},
\eeq
where  $\{\balpha \}$  is the set of the six roots of  $SL(3)$. In terms of
fermions or bosons the screening currents look like
\bea\label{3.2.15}
J^{(\bsalpha _1)} &= \psi ^{(1)\ast }\psi ^{(2)\ast } = \exp (\phi _1 + \phi _2
)
\nn\\
J^{(\bsalpha _2)} &= \psi ^{(2)\ast }\psi ^{(3)\ast } = \exp (\phi _2 + \phi _3
)
\nn\\
J^{(\bsalpha _3)} &= \psi ^{(1)}\psi ^{(3)\ast } = \exp (\phi _3 - \phi _1)
\nn\\
J^{(-\bsalpha _1)} &= \psi ^{(1)}\psi ^{(2)} = \exp (- \phi _1 - \phi _2)
\nn\\
J^{(-\bsalpha _2)} &= \psi ^{(2)}\psi ^{(3)} = \exp (- \phi _2 - \phi _3)
\nn\\
J^{(-\bsalpha _3)} &= \psi ^{(3)}\psi ^{(1)\ast } = \exp (\phi _1 - \phi _3)
\eea
The particular $\tau $-function is now described in terms of the correlator
\beq\label{3.2.16}
\tau ^{(3)}_\bsN(\bx) = \langle \bN|e^{H(\bsx)}G|0\rangle
\eeq
with
\beq\label{3.2.17}
G \sim  \prod  _\bsalpha \exp Q^{(\bsalpha )}
\eeq
The condition of Cartan neutrality is preserved by compensation of charges
between the operator (\ref{3.2.17}) and left vacuum
  $\langle \bN|$  in (\ref{3.2.16}). It
is obvious that in such case due to the condition of Cartan neutrality of the
correlator (like in Wess-Zumino models) the mode  $\tilde J =
\partial \tilde \phi  = {\displaystyle1\over
\displaystyle \sqrt{p}}\sum ^p_{i=1}\partial \phi _i$
decouples from the correlator, and
\bea\label{3.2.18}
\tau^{(3)}_\bsN(\bx)
&= \langle \bN|e^{H(\bsx)}G|0\rangle  =\nn\\
&= \left.\langle 0|\exp \left( \sum_{k>0}
\tilde J_k \sum^3_{i=1} x^{(i)}_k \right)
|0\rangle_{\tilde\phi}
 \langle \bN|e^{H(t,\bar t)}G|0\rangle\right|_{\sum_i\phi_i=0}
\eea
where the first correlator in the second row is trivially equal to unity. For
the specific choice of the operator  $G$  in (\ref{3.2.18})
\beq\label{3.2.19}
G = G_{1,2} = \exp \left( \int   J^{(\bsalpha _1)}\right) \exp \left( \int
J^{(\bsalpha _2)}\right)
\eeq
we reproduce the formula (\ref{zsl3}).

Finally, let us only stress two main ideas we have demonstrated above:
first, we proved that the solutions to the discrete Virasoro-$W$
constraints can be rewritten from (free) bosonic to (free) fermionic
language, which means {\it automatically} that they are solutions to
integrable systems in the sense of \cite{DJKM}. Second, in general case
the solutions to discrete constraints are presented in the form of
conformal multimatrix models, being particular solutions to {\it
multi}component hierarchies.

\section{A solution to the continuum Virasoro -$W$ constraints}

\bigskip
This section is devoted to the derivation of the solution to the
continuous Virasoro and  ${\cal W}$- constraints. To be more precise,
we shall investigate them as being the direct consequence of the
''matrix" equations, which could be treated later on as the Ward
identites for certain matrix integrals.

This is however not exactly the same what we had before for the case of
discrete constraints. The reason is that the continuum case differs from
the discrete one roughly speaking by replacement of ordinary
free scalar fields by the same fields but with {\it antiperiodic}
boundary conditions. This is the sense of so-called {\it double scaling
limit} (see below for details) which can be done successfully for all
conformal multimatrix models \cite{KMMMP,MP}, discussed above.
The exact solution in terms of
conformal correlators is much more complicated for the antiperiodic
fields
\footnote{ one of the reasons is absense of a rather simplifying
selection rule with a zero-mode (\ref{discrG}) },
thus instead here we are going to reformulate the problem.

Fortunately, it turns out that the continuum Virasoro constraints can be
summed up into certain {\it matrix} differential operators. Namely, for
the $W^{(p)}$-algebra these operators are related to the Laplacians (or
better Casimirs) for corresponding algebras, having the form of
\beq\label{casim}
{\partial ^p \over \partial \Lambda ^p} + ...
\eeq
where $\Lambda$ is $N \times N$ hermitean matrix (for the $SU(N)$-case).
These equations might be identified with the Ward identities for
''continuum" matrix theories.

So, we start with an equation of the type of (\ref{casim}) operator
vanish certain function, and prove that it is equivalent to the
continuum Virasoro (or $W$) constraints.
\beq\label{wi}
\{Tr\ \epsilon (\Lambda )[V\ '(\partial /\partial \Lambda _{tr}) -
\Lambda ]\} {\cal F}[\Lambda ] = 0\hbox{.}
\eeq
(\ref{wi}) can be certainly interpreted as a ward identity
satisfied by a matrix integral which after the proper normalization
gives a solution to continuum $2d$ gravity. The exact formula for
corresponding partition function reads
\beq\label{gkm}
Z^{(N)}[V|M] \equiv  C^{(N)}[V|M] e^{TrV(M)-TrMV'(M)}\int
DX\ e^{-TrV(X)+TrV'(M)X}
\eeq
where the integral is taken over  $N\times N$  ``Hermitean" matrices,
with the normalizing factor given by
Gaussian integral
\beq\label{norm}
C^{(N)}[V|M]^{-1} \equiv  \int \hbox{  DY }\ e^{-TrU_2[M,Y]},
$$
$$
U_2 \equiv \lim _{\epsilon \rightarrow 0}
{1\over \epsilon ^2}Tr[V(M+\epsilon Y) - V(M) - \epsilon YV'(M)]
\eeq
and discuss only specific potentials,  $V (X) =
const\cdot X^{p+1}$, giving rise (while substituted in (\ref{wi}))
to (\ref{casim}).
\footnote{The proof of the Virasoro constraints for
generic potential is based on integrability and will be presented
below.}

In the simplest example of $p=2$ we have quadratic operator (or just
Laplacian) and will prove the identity

\beq\label{vir}
{1\over {\cal F}} tr(\epsilon {\partial ^2\over \partial \Lambda ^2_{tr}}
 - \epsilon \Lambda ){\cal F} = {1\over Z}
\sum _{n\geq -1} {\cal L}_nZ \hbox{   } tr(\epsilon \Lambda ^{-n-2})
\eeq
for

\beq\label{F}
{\cal F}^{\{2\}}\{\Lambda \} \equiv  \int   DX\ \exp (- trX^3/3 + tr\Lambda X)
= C[\sqrt{\Lambda }] \exp ({2\over 3}tr\Lambda ^{3/2})
Z^{\{2\}}(T_m)
$$
$$
T_m = {1\over m}Tr M^{-m} = {1\over m} Tr \Lambda ^{-m/2},
\ \ \ m\ \ -\ \ odd
\eeq
with

\beq\label{cl}
C[\sqrt{\Lambda }] = {\rm det} (\sqrt{\Lambda }\otimes I +
I\otimes \sqrt{\Lambda })^{-{1\over2}}
\eeq
and

\bea\label{l2}
{\cal L} _n = {1\over 2}\sum _{{k>\delta_{n+1,0} }\atop {k\ odd}}
kT_k {\partial \over \partial T_{k+2n}} + {1\over 4}
\sum _{^{a+b=2n}_{a,b> 0\ ;\ a,b\ odd}}
{\partial ^2\over \partial T_a\partial T_b} + \nn \\
+ \delta _{n+1,0}\cdot {T^2_1\over 4} + \delta _{n,0}\cdot {1\over 16}
- {\partial \over \partial T_{2n+3}}\hbox{ . }
\eea

  While (\ref{vir})
is valid  for {\it any size} of the matrix $\Lambda $, in the limit of
infinitely large $\Lambda $ ($N \to \infty$) we can insist that
all the quantities

\beq
tr(\epsilon \Lambda ^{-n-2})
\eeq
$(e.g$. $tr\Lambda ^{p-n-2}$ for $\epsilon  = \Lambda ^p)$
become algebraically independent, so that eq.
(\ref{vir}) implies that

\beq
{\cal L} _nZ\{T\} = 0\hbox{, }    n \geq  -1 \ .
\eeq

\bigskip
Note that ${\cal F} \{\Lambda \}$ in (\ref{F}), which
we have to differentiate in order to prove (\ref{vir}),
depends only upon eigenvalues
$\{\lambda _k\}$ of the matrix $\Lambda $. Therefore, it is natural to consider
eq.(\ref{vir}) at the diagonal point $\Lambda _{ij}=0$, $i\neq j$. The only
``non-diagonal" piece of (\ref{vir}) which survives at this point is
proportional to

\beq\label{llm}
\left.{\partial ^2\lambda _k\over \partial \Lambda _{ij}\partial \Lambda _{ji}}
\right| _{\Lambda _{mn}=0\hbox{, } m\neq n} =
{\delta _{ki}-\delta _{kj}\over \lambda _i-\lambda _j}\hbox{
for }\ i\neq j.
\eeq
Eq.(\ref{llm}) is nothing but a familiar formula for the second order
correction to the Hamiltonian
eigenvalues in ordinary quantum-mechanical perturbation theory. It can be
easily derived from the variation of determinant formula:

\begin{eqnarray}
\delta log({\rm det} \ \Lambda ) = tr {1\over \Lambda } \delta \Lambda  -
{1\over 2}
tr( {1\over \Lambda } \delta \Lambda  {1\over \Lambda } \delta \Lambda )
 + \ldots \hbox{ .}
\end{eqnarray}
For diagonal $\Lambda _{ij}= \lambda _{i}\delta _{ij}$, but, generically,
non-diagonal $\delta \Lambda _{ij}$, this equation gives

\begin{eqnarray*}
\sum  _k {\delta \lambda _k\over \lambda _k} = - {1\over 2} \sum _{i\neq j}
{\delta \Lambda _{ij}\delta \Lambda _{ji}\over \lambda _i\lambda _j} =
{1\over 2} \sum _{i\neq j} \left( {1\over \lambda _i} -
{1\over \lambda _j}\right)
{\delta \Lambda _{ij}\delta \Lambda _{ji}\over \lambda _i-\lambda _j} +
\ldots \hbox{ ,}
\end{eqnarray*}
which proves (\ref{llm}).

Since {\it matrix} $\epsilon $ is assumed
to be a function of $\Lambda $, it can be, in fact, treated as a function of
eigenvalues $\lambda _i$. Then, we use actually only\\
a) the concrete form of the normalization (\ref{norm})\\
b) the fact that $Z[T(\lambda _i)]$ is a {\it complicated} function,
$i.e.$ we should differentiate it as depending on $\{\lambda _i\}$ only
through time variables. After that, (\ref{vir}) can be rewritten in the
following way:

\beq
{e^{-{2\over 3}tr\Lambda ^{3/2}}\over C(\sqrt \Lambda )Z\{T\}}\left[ tr\
\epsilon \{{\partial ^2\over \partial \Lambda ^2} -
\Lambda \}\right]  C(\sqrt \Lambda ) e^{
{2\over 3}tr\Lambda ^{3/2}}Z\{T\} =
\eeq

\beq\label{pr1}
= {1\over Z}\sum _{a,b> 0}
{\partial ^2Z\over \partial T_a\partial T_b}
\sum  _i \epsilon (\lambda _i)
{\partial T_a\over \partial \lambda _i}\cdot {\partial T_b\over
\partial \lambda _i} +
\eeq

\bea\label{pr2}
+ {1\over Z} \sum _{n\geq 0}
{\partial Z\over \partial T_n} \left[ \sum _{i,j}
\epsilon (\lambda _i)
{\partial ^2T_n\over \partial \Lambda _{ij}\partial \Lambda _{ji}} + 2 \sum  _i
\epsilon (\lambda _i) {\partial T_n\over \partial \lambda _i}
{\partial logC\over \partial \lambda _i} +\right. \nn \\
+ \left. 2 \sum  _i \epsilon (\lambda _i) {\partial T_n\over \partial
\lambda _i}
\left( {2\over 3}\right)  {\partial \over \partial \lambda _i}
tr\Lambda ^{3/2} \right]  +
\eea

\beq\label{pr3}
+ \left[
 \sum  _i
\epsilon (\lambda _i)
\left( {\partial \over \partial \lambda _i}\left(
{2\over 3}\right)  tr\Lambda ^{3/2}\right) ^2 \right.
-  \sum  _i \lambda _i \epsilon (\lambda _i) +
\eeq

\beq\label{pr4}
+ \sum _{i,j}\epsilon (\lambda _i)
\left( {\partial ^2\over \partial \Lambda _{ij}\partial \Lambda _{ji}}
\left( {2\over 3}\right)  tr\Lambda ^{3/2}\right)  +
\eeq

\beq\label{pr5}
+ 2 \sum  _i \epsilon (\lambda _i) \left(
{2\over 3}\right)
{\partial tr\Lambda ^{3/2}\over \partial \lambda _i}
{\partial logC\over \partial \lambda _i} +
\eeq

\beq\label{pr6}
+ \left. {1\over C} \sum _{i,j}\epsilon (\lambda _i)
{\partial ^2C\over \partial \Lambda _{ij}\partial \Lambda _{ji}} \right]
\eeq
with  $tr\Lambda ^{3/2} = \sum  _k \lambda ^{3/2}_k$ and

\beq\label{c1}
C = \prod _{i,j}(\sqrt{\lambda }_i + \sqrt{\lambda }_j)^{-1/2}.
\eeq

The calculation of all the quantities in (\ref{pr1}) - (\ref{pr6}) is
just an exercise of taking derivatives, using (\ref{llm}), all necessary
details can be found in \cite{MMM91b,KMMMZ91b}. Careful calculation
shows that all the terms after taking the derivatives contain only {\it
negative} powers of $\sqrt \lambda _i$ and can be ''arbsorbed" in times.
The result is:

\bea
{e^{-{2\over 3}tr\Lambda ^{3/2}}\over C(\sqrt \Lambda )Z\{T\}}\left[ tr\
\epsilon \{{\partial ^2\over \partial \Lambda ^2} -
\Lambda \}\right]  C(\sqrt \Lambda ) e^{
{2\over 3}tr\Lambda ^{3/2}}Z\{T\} = \nn \\
= {1\over Z} \sum _{n\geq -1}tr(\epsilon _p\Lambda ^{-n-2})
\left\lbrace {1\over 2}\sum _{k>\delta _{n+1,0}}kT_k
{\partial \over \partial T_{2n+k}}
+{1\over 4}\sum _{^{a+b=2n}_{a>0,b>0}}{\partial ^2\over \partial T_a
\partial T_b} +\right.\nn \\
+ \left.
{1\over 16} \delta _{n,0} + {1\over 4} \delta _{n+1,0} T^2_1 - {\partial
\over \partial T_{2n+3}}
\right\rbrace  Z(T) = 0.
\eea
or just the set of Virasoro constraints for
the case $p=2$.

In the case of generic  $p$  the analog of the derivation
actually involves the same steps:

-- Represent  ${\cal F} [\Lambda ]$  as

\beq
{\cal F} ^{\{p\}}[\Lambda ] = g_p[\Lambda ]Z^{\{p\}}(T_n)\hbox{,}
\eeq
with

\beq
g_p[\Lambda ] = {\Delta (M)\over \Delta (\Lambda )}\prod _i
[V''(\mu _i)^{-1/2} e^{(\mu _i V'(\mu _i)-V (\mu _i))}]
= {\Delta (\Lambda ^{1/p})\over \Delta (\Lambda )}\prod _i
[\lambda ^{-{p-1\over 2p}}_ie^{ {p\over p+1}\lambda ^{1+1/p}_i}]\hbox{.}
\eeq

-- Substitute this  ${\cal F} ^{\{p\}}[\Lambda ]$  into (\ref{wi}), which in
the
particular case of  $\displaystyle {V _p(X) = {X^{p+1}\over {p+1}}}$,
looks like

\beq\label{wip}
\{Tr\ \epsilon (\Lambda )[({\partial \over \partial \Lambda _{tr}})^p -
\Lambda ]\} g_p[\Lambda ]Z^{\{p\}}(T_n) = 0.
\eeq
Higher-order derivatives,  $\displaystyle {{\partial ^iZ\over
\partial \Lambda ^i_{tr}}}$ ,
they are defined with the help of relations like (\ref{llm}).

-- Perform the shift of variables

\beq
T_n\to \hat T_n = T_n -
{p\over n}\delta _{n,p+1}
\eeq
(this procedure doesn't change the derivatives).

-- After all these substitutions the $l.h.s$. of eq.(\ref{wip}) acquires the
form of an infinite series where every item is a product of
$Tr[\tilde \epsilon (M)M^{-k}]$  and a linear combination of generators of
${\cal W}_p$-algebra,
acting on  $Z^{\{p\}}(T_n)$. In the case of  $p = 3$  this
equation looks like

\beq\label{w}
\new
\begin{array}{c}
{1\over 27}
Tr\left[\tilde \epsilon (M)M^{-3}\left\lbrace \sum M^{-3n}
{\cal W} ^{(3)}_{3n} +\right.\right. \\
+ 9\sum M^{-3n-1/3} \left\{ \sum  (3k-2)\hat T_{3k-2}
{\cal W} ^{(2)}_{3n+3k} +
\sum  {\partial \over \partial T_{3a+1}}
{\cal W} ^{(2)}_{3b-3}\right\}+ \\
\left. \left. + 9\sum M^{-3n-2/3}\left\{ \sum  (3k-2)\hat T_{3k-2}
{\cal W} ^{(2)}_{3n+3k} +
\sum  {\partial \over \partial T_{3a+1}}
{\cal W} ^{(2)}_{3b-3}\right\}  \right\rbrace \right] Z^{\{3\}} = 0,
\end{array}
\eeq

-- If  $N = \infty $  all the quantities  $Tr\tilde \epsilon (M)M^{-k}$ with
given  $k$  but varying  $\tilde \epsilon (M)$ become independent, and
(\ref{wip})
may be said to give $W$-constraints. The exact proof for the case $p=3$
can be found in \cite{Mikh92}.

\section{Integrability of GKM}

The purpose of this section is to prove that the solution to the
continuum Virasoro- and $W$- constraints found above is nothing but a
particular solution to the integrable KP system. Namely:

 ($A$) The partition function $Z^V_N[M]$ (\ref{gkm}),
if considered as a function of time-variables \cite{Miwa}

\beq\label{miwa}
T_k = {1\over k} Tr\ M^{-k}\hbox{, }  k\geq 1\hbox{  ;}
\eeq
is a KP $\tau$-function for {\it any} value of $N$ and {\it any}
potential $ V [X]$.

($B$) As soon as  ${ V}[X]$  is homogeneous polynomial of degree
$p+1$,
$Z^ {\{V\}}_N[M] = Z^{\{p\}}_N[M]$  is in fact a $\tau $-function of
$p$-reduced KP hierarchy \cite{SW85}.
\footnote{Moreover, actually,
$\displaystyle {{\partial Z^{\{p\}}\over \partial T_{np}}}=0$.}

In order to prove these statements, first,  we rewrite (\ref{gkm})
in terms of determinant formula

\beq\label{det}
Z^ {\{V\}}_N[M] = {{\rm det} _{(ij)}\phi _i(\mu _j)\over \Delta
(\mu )}\
\ \ \ \ i,j =
1,...,N.
\eeq
Then, we show that {\it any} KP $\tau $-function in the Miwa
parameterization does have the same determinant form.
\footnote{As a check of self-consistency, it can be proven (see for
example \cite{KMMMZ91b})
that any determinant formula (\ref{det}) with {\it any} set of functions
$\{\phi _i(\mu )\}$  satisfies the Hirota difference bilinear equation.}

The main thing which distinguishes matrix models from the point of view of
solutions to the KP-hierarchy is that the set of functions  $\{\phi _i(\mu )\}$
in (\ref{det}) is not arbitrary. Moreover, this whole {\it infinite} set of
functions is
expressed in terms of a {\it single} potential  ${ V}[X]$ ($i.e$. instead
of
arbitrary {\it matrix  $A_{ij}$} in  $\phi _i(\mu ) = \sum A_{ij}\mu ^j$
we have here
only a {\it vector}  $ V_i$ or  $ V[\mu ] =
\sum  V_i\mu ^i)$. This is
the origin of  ${\cal L}_{-1}$ and other  ${\cal W}$- constraints (which in the
context of KP-hierarchy may be considered as implications of  ${\cal L}_{-1})$.
All these constraints are in fact
contained in the Ward identity (\ref{wi}).

\subsection{Integrability from the determinant formula}

\bigskip
We begin from evaluation of the integral:

\beq
{\cal F}^{\{V\}}_N[\Lambda ] \equiv  \int   DX\ e^{- Tr[ V(X)
-
Tr\Lambda X]}\hbox{.}
\eeq
The integral over the ''angle" $U(N)$-matrices can be easily taken
\cite{IZ,Mehta}
and if eigenvalues of  $X$  and  $\Lambda $  are denoted by  $\{x_i\}$  and
$\{\lambda _i\}$  respectively, this integral can be rewritten as

\beq\label{int}
{1\over \Delta (\Lambda )}\left[  \prod _{i=1}^N
\int   dx_ie^{- V(x_i)+\lambda _ix_i} \right] \Delta (X)\hbox{.}
\eeq
$\Delta (X)$  and  $\Delta (\Lambda )$  are Van-der-Monde
determinants, $e.g$. $\Delta (X) = \prod_{i>j}(x_i-x_j)$.

The $r.h.s$. of (\ref{int}) can be rewritten as

\beq
\Delta ^{-1}(\Lambda) \Delta ({\partial \over \partial \Lambda})
\prod _i \int dx_i e^ {- V(x_i) + \lambda _i x_i } =
$$
$$
=
\Delta ^{-1}(\Lambda ) \hbox{det} _{(ij)}F_i(\lambda _j)
\eeq
with

\beq
F_{i+1}(\lambda ) \equiv  \int   dx\ x^ie^{- V(x)+\lambda x} =
({\partial \over \partial \lambda })^iF_1(\lambda ).
\eeq
Note that

\beq
F_1(\lambda ) = {\cal F}^{\{{V}\}}_{N=1}[\lambda ]\hbox{ . }
\eeq
If we recall that

\beq
\Lambda  = V'(M)
\eeq
and denote the eigenvalues of  $M$  through  $\{\mu _i\}$ , then:

\beq
{\cal F}^{\{V\}}_N[ V'(M)] = {{\rm det} \ \tilde
\Phi _i(\mu _j)\over
\prod _{i>j}(V'(\mu _i)- V'(\mu _j))}\ ,
\eeq
with

\beq\label{entry1}
\tilde \Phi _i(\mu ) = F_i(V'(\mu )).
\eeq

\bigskip
Proceed now to the normalization (\ref{norm}). Indeed, it is given by
the Gaussian integral:

\beq\label{norm1}
C^{(N)}[V|M]^{-1} \equiv  \int   DX\ e^{-U_2(M,X)}.
\eeq
Making use of $U(N)$-invariance of Haar measure $dX$ one can easily diagonalize
$M$. Of course, this does not imply any integration over angular
variables and provide no factors like  $\Delta (X)$. Then for evaluation of
(\ref{norm1}) it remains to use the obvious rule of Gaussian integration,

\beq
\int   DX\ e^{-\sum ^N_{i,j} U_{ij}X_{ij}X_{ji}} \sim \prod ^N_{i,j}
U^{-1/2}_{ij}
\eeq
and substitute the explicit
expression for $U_{ij}(M)$. If potential is represented as a formal series,

\beq
V(X) =\sum  {v_n\over n}X^n,
\eeq
we have

$$
U_2(M,X)
=\sum ^\infty _{n=0}v_{n+1}\left\lbrace \sum _{a+b=n-1}TrM^aXM^bX
\right\rbrace ,
$$
and

$$
U_{ij} =\sum ^\infty _{n=0}v_{n+1} \left\lbrace
\sum _{a+b=n-1}\mu ^a_i\mu ^b_j \right\rbrace  =
\sum ^\infty _{n=0}v_{n+1} {\mu ^n_i - \mu ^n_j\over \mu _i -
\mu _j}  = {V'(\mu _i) -  V'(\mu _j)\over \mu _i - \mu _j}.
$$
Coming back to (\ref{gkm}), we conclude that

\bea\label{**}
Z^{\{V\}}_N[M] = e^{Tr[V(M)-MV'(M)]}
C^{(N)}[V|M] {\cal F}_N[V'(M)] \sim \nn \\
\sim [{\rm det} \ \tilde \Phi _i(\mu _j)] \prod _{i>j}^N
{U_{ij}\over (V'(\mu _i)-V'(\mu _j))} \prod _{i=1}
s(\mu _i)  =  {[{\rm det} \ \tilde \Phi _i(\mu _j)]\over \Delta (M)}
\prod _{i=1}^N
s(\mu _i)\ .
\eea

\beq\label{entry2}
s(\mu ) = [ V''(\mu )]^{1/2} e^{V(\mu )-\mu  V'(\mu )}
\eeq
The product of $s$-factors at the $r.h.s$. of (\ref{**}) can be absorbed into
$\tilde \Phi $-functions:

\beq\label{zvdet}
Z^{\{ V\}}_N[M] = {{\rm det}
\Phi _i(\mu _j)\over \Delta (M)}\hbox{,}
\eeq
where

\beq\label{entry3}
\Phi _i(\mu ) = s(\mu )\tilde \Phi _i(\mu )
\stackreb{\mu \to
\infty}{\to} \mu ^{i-1}(1 + {\cal O}({1\over \mu})).
\eeq
where the asymptotic is crucial for the determinant (\ref{zvdet}) to be
a solution to the KP hierarchy in the sense of \cite{SW85}.

\bigskip
{\it The Kac-Schwarz operator} \cite{KSch,Sch}.
{}From eqs.(\ref{entry1}),(\ref{entry2}) and (\ref{entry3}) one can deduce that
$\Phi _i(\mu )$
can be derived from the basic function  $\Phi _1(\mu )$  by the relation

\beq
\Phi _i(\mu ) = [V''(\mu )]^{1/2}\int   x^{i-1}
e^{- V (x) + x  V ' (\mu )}dx =
A_{\{ V\}}^{i-1}(\mu )\Phi _1(\mu )\ ,
\eeq
where  $A_{\{ V\}}(\mu )$  is
the first-order differential operator

\bea\label{ks}
A_{\{ V\}}(\mu ) & = & s {\partial \over \partial \lambda } s^{-1} =
{e^{ V(\mu )-\mu  V'(m)}\over [V''(\mu )]^{1/2}}
{\partial \over \partial \mu }
{e^{- V(\mu )+\mu  V'(\mu )}\over [ V''(\mu )]^{1/2}} =
\nn \\
& = & {1\over  V''(\mu )} {\partial \over \partial \mu } + \mu  -
{ V'''(\mu )\over 2[ V''(\mu )]^2}\ .
\eea
In the particular case of $V(x) = {{x^{p+1}} \over {p+1}}$

$$
A_{\{p\}}(\mu ) =
{1\over p\mu ^{p-1}} {\partial \over \partial \mu } + \mu  -
{p-1\over 2p\mu ^p}
$$
coincides (up to the scale transformation of  $\mu $  and
$A_{\{p\}}(\mu ) )$ with the operator which determines the finite dimensional
subspace
of the Grassmannian in ref.\cite{KSch}
We emphasize that the property

\beq
\Phi _{i+1}(\mu ) = A_{\{V\}}(\mu ) \Phi _i(\mu )\ \ \ \
( F _{i+1}(\lambda )
= {\partial \over \partial \lambda } F _i(\lambda ) )
\eeq
is exactly the thing which distinguishes partition functions of GKM from
the expression for generic $\tau $-function in Miwa's coordinates,

\beq\label{taumiwa}
\tau ^{\{ \phi _i\}}_N[M] = {[{\rm det} \ \phi _i(\mu _j)]\over \Delta (M)}
\hbox{,}
\eeq
with arbitrary sets of functions  $\phi _i(\mu )$. In the next section we
demonstrate that the quantity (\ref{taumiwa}) is exactly a KP $\tau $-function
in Miwa coordinates.

\subsection{KP $\tau $-function in Miwa parameterization}

A generic KP $\tau $-function is a correlator of a special form
\cite{DJKM}:

\beq
\tau ^G\{T_n\} = \langle 0|:e^{\sum \ T_nJ_n}: G|0\rangle
\eeq
with

\beq\label{ff}
J(z) = \tilde \psi (z)\psi (z)\hbox{; }  G =\
:\exp \ {\cal G}_{mn}\tilde \psi _m\psi _n:
\eeq
in the theory of free 2-dimensional fermionic fields  $\psi (z)$,
$\tilde \psi (z)$  with the action $\int
\tilde \psi \bar \partial \psi $. The vacuum states are defined by conditions

\beq
\psi _n|0\rangle  = 0\ \ n < 0\hbox{ , }  \tilde \psi _n|0\rangle  = 0\ \ n
\geq  0
\eeq
where  $\psi (z) =
\sum _{\bf Z}
\psi _nz^n\ dz^{1/2} $ , $\tilde \psi (z) =
\sum _{\bf Z}
\tilde \psi _nz^{-n-1}\ dz^{1/2}$.

The crucial restriction on the form of the correlator, implied by
(\ref{ff}) is
that the operator  $:e^{\sum \ T_nJ_n}:$  $G$  is {\it Gaussian} exponential,
so that the insertion of this operator may be considered just as a
modification of  $\langle \tilde \psi \psi \rangle$  {\it propagator}, and
the Wick theorem is applicable. Namely, the correlators

\beq\label{wick1}
\langle 0| \prod _i
\tilde \psi (\mu _i)\psi (\lambda _i) G|0\rangle
\eeq
for {\it any} relevant $G$ are expressed through the pair correlators of the
same form:

\beq\label{wick2}
(\ref{wick1}) =
{\rm det} _{(ij)} \langle 0| \tilde \psi (\mu _i) \psi (\lambda _j)
G|0\rangle
\eeq

The simplest way to understand what happens to the operator  $e^{\sum T_nJ_n}$
after the substitution of (\ref{miwa})
is to use the free-{\it boson}
representation
of the current  $J(z)=\partial \varphi (z)$. Then  $\sum  T_nJ_n =
\displaystyle {\sum _i
\left\lbrace  \sum _n {1\over n\cdot \mu _i^n}
\varphi _n\right\rbrace}  = \sum _i \varphi (\mu _i)$, and

\beq
:e^{\sum _i\varphi (\mu _i)}: = {1\over \prod _{i<j}(\mu _i-\mu _j)}
\prod _i :e^{\varphi (\mu _i)}:\hbox{ .}
\eeq
In fermionic representation it is better to start from

\beq
T_n = {1\over n} \sum _i ({1\over \mu ^n_i} - {1\over \tilde \mu _i^n} )
\eeq
instead of (\ref{miwa}). Then

\beq
:e^{\sum T_nJ_n}: = {\prod _{i,j}^N(\tilde \mu _i-\mu _j)\over
\prod _{i>j}(\mu _i-\mu _j) \prod _{i>j}(\tilde \mu _i-\tilde \mu _j)}
\prod _i \tilde \psi (\tilde \mu _i)\psi (\mu _i)\hbox{ .}
\eeq
In order to come back to (\ref{miwa}) it is necessary to shift all
$\tilde \mu _i$'s  to infinity. This may be expressed by saying that the left
vacuum is substituted by

$$
\langle N| \sim
\langle 0|\tilde \psi (\infty )\tilde \psi '(\infty )...\tilde \psi ^{(N-1)}(
\infty ).
$$
The $\tau $-function now can be represented in the form:

\beq
\new
\begin{array}{l}
\tau ^G_N[M] = \langle 0|:e^{\sum T_nJ_n}:G|0\rangle  =
\Delta (M)^{-1}\langle N|\prod _i
:e^{\varphi (\mu _i)}: G|0\rangle  = \\
=\lim _{\tilde \mu _j \to \infty}{\prod _{i,j}(\tilde \mu _i-\mu _j)
\over \prod _{i>j}
(\mu _i-\mu _j) \prod _{i>j}(\tilde \mu _i-\tilde \mu _j)}\langle 0|\prod _i
\tilde \psi (\tilde \mu _i)\psi (\mu _i)G|0\rangle
\end{array}
\eeq
applying the Wick's theorem (\ref{wick1}), (\ref{wick2}) and taking the
limit $\tilde \mu _i \to \infty$ we obtain:

\beq
\tau ^G_N[M] =
{{{\rm det} \ \phi _i(\mu _j)}\over {\Delta (M)}}
\eeq
with functions

\beq\label{2point}
\phi _i(\mu ) \sim  \langle 0|\tilde \psi ^{(i-1)}(\infty )\psi (\mu )
G|0\rangle
\stackreb{\mu \to
\infty}{\to} \mu ^{i-1}(1 + {\cal O}({1\over \mu})).
\eeq
Thus, we proved that KP $\tau $-function in Miwa coordinates (\ref{miwa}) has
exactly the determinant form (\ref{det}), or, put differently,
(\ref{det}) is a $\tau $-function of KP hierarchy. Below we will discuss
how from generic point of Grassmannian described by $G = \exp {\sum
A_{ij}\tilde \psi _i \psi _j}$ or infinite matrix with {\it two} indices
($\infty ^2$) one can restrict it to a solution, determined only by one
function ($\infty $) or two functions ($2 \times \infty$)..

\section{Universal ${\cal L}_{-1}$-constraint and string equation}

Let us return to the question of specifying particular ''stringy"
solutions to the KP hierarchy. It turns out that using integrability it
is enough to prove {\it only} the so-called {\it string equation} or
${\cal L}_1$-constraint, all other recursion relation follow from these
two statements \cite{FKN1,FKN2}.

It is well known that  ${\cal L}_{-1}$-constraint is
closely related to the action of operator

\beq\label{trlam}
Tr{\partial \over \partial \Lambda _{tr}} = Tr {1\over  V''(M)}
{\partial \over \partial M_t}_r.
\eeq
Therefore it is natural to examine, how this operator acts on

\beq\label{z}
Z^{\{ V\}}[M] = {{\rm det} \
\tilde \Phi _i(\mu _j)\over \Delta (M)}\prod _i
s(\mu _i),
\eeq

\beq
s(\mu ) = ( V''(\mu ))^{1/2}e^{ V(\mu )-\mu  V'(\mu )},
\eeq

$$
\tilde \Phi _i(\mu ) = F_i(\lambda ) =
(\partial /\partial \lambda )^{i-1}F_1(\lambda )\hbox{, }   \lambda  =
V'(\mu ).
$$
First of all, if $Z^{\{ V\}}$ is considered as a function
of  $T$-variables,

\beq
{1\over Z^{\{ V\}}}Tr{\partial \over \partial \Lambda _{tr}}
Z^{\{ V\}} = -\sum _{n\geq 1}Tr [{1\over  V''(M)M^{n+1}}]
{\partial logZ^{\{ V\}}\over \partial T_n}.
\eeq

On the other hand, if we apply (\ref{trlam}) to explicit formula
(\ref{z}), we obtain:

\beq
\new
\begin{array}{c}
{1\over Z^{\{V\}}}Tr{\partial \over \partial \Lambda _{tr}}
Z^{\{ V\}}\\
 = - Tr\ M + {1\over 2} \sum _{i,j}{1\over
 V''(\mu _i) V''(\mu _j)}
{ V''(\mu _i)- V''(\mu _j)\over \mu _i - \mu _j} +
Tr{\partial \over \partial \Lambda _{tr}}\log \ {\rm det} \ F_i(\lambda _j),
\end{array}
\eeq

Below, we will argue that

\beq\label{se}
{{1\over Z^{\{ V\}}}{\cal L}^{\{ V\}}_{-1}Z^{\{ V\}}} = -
{\partial \over \partial T_1}\log \ Z^{\{ V\}} + TrM -
Tr{\partial \over \partial \Lambda _{tr}}\log \ {\rm det} \ F_i(\lambda _j).
\eeq
can be used in order to suggest the formula for the
universal operator  ${\cal L}^{\{ V\}}_{-1}$.

Here

\bea\label{l-1}
{\cal L}^{\{V \}}_{-1} =\sum _{n\geq 1}Tr
[{1\over  V''(M)M^{n+1}}] {\partial \over \partial T_n} + \nn \\
+ {1\over 2}
\sum _{i,j}{1\over V''(\mu _i) V''(\mu _j)}{ V''
(\mu _i)-
 V''(\mu _j)\over \mu _i - \mu _j} - {\partial \over \partial T_1},
\eea
and this turns into more common expression when
$ V(X) = X^{p+1}/(p+1)$  (note that the
items with  $i=j$  are included into the sum at the $r.h.s$. in
(\ref{l-1})).

So, in order to prove the ${\cal L}^{\{ V \}}_{-1}$-constraint, one
should prove that the $r.h.s$. of (\ref{se}) vanishes, i.e.

\beq\label{dt1}
{\partial \over \partial T_1}\log \ Z^{\{ V\}}_N = TrM -
Tr{\partial \over \partial \Lambda _{tr}}\log \ {\rm det} \ F_i(\lambda _j)
\hbox{,}
\eeq
This is possible to prove only if we remember that
$Z^{\{ V\}}_N = \tau ^{\{ V\}}_N$. In this case the
$l.h.s$. may be represented as residue of the ratio

\beq\label{res}
res_\mu {\tau ^{\{ V\}}_N(T_n+
\mu ^{-n}/n)\over \tau ^{\{ V\}}_N(T_n)} =
{\partial \over \partial T_1}\log \ \tau ^{\{ V\}}_N(T_n).
\eeq
However, if expressed through Miwa coordinates, the $\tau $-function in the
numerator is given by the same formula with one {\it extra} parameter  $\mu $ ,
$i.e$. is in fact equal to $\tau ^{\{ V\}}_{N+1}$ . This idea is almost
enough to deduce (\ref{dt1}). Let us begin with an illustrative example of
$N = 1$. Then $(\lambda  = V'(\mu ))$

$$
\tau ^{\{ V\}}_1(T_n) = \tau ^{\{ V\}}_1[\mu _1] =
e^{ V(\mu _1)-\mu _1 V'(\mu _1)}[{ V''}(\mu _1)]^{1/2}F(
\lambda _1){ , }
$$

\beq\label{315}
\new
\begin{array}{c}
\tau ^{\{ V\}}_1(T_n+\mu ^{-n}/n) =
\tau ^{\{ V\}}_2[\mu _1,\mu ] = \\
= e^{ V(\mu _1)-\mu _1 V'(\mu _1)}e^{ V(\mu )-\mu
 V'(
\mu )} {[ V''(\mu _1) V''(\mu )]^{1/2}\over \mu  -
\mu _1}[F(\lambda _1)\partial F(\lambda )/\partial \lambda  -
F(\lambda )\partial F(\lambda _1)/\partial \lambda _1] = \\
= {e^{V(\mu )-\mu  V'(\mu )}[ V''(\mu )]^{1/2}
F(\lambda )\over \mu  - \mu _1} \tau ^{\{ V\}}_1[\mu _1]\cdot [ -
\partial logF(\lambda _1)/\partial \lambda _1 +
\partial logF(\lambda )/\partial \lambda ].
\end{array}
\eeq
The function

\beq
F(\lambda ) =\int   dx\ e^{- V(x)+\lambda x} \sim
e^{ V(\mu )-\mu  V'(\mu )} [ V''(\mu )]^{-1/2}\{1 +
O({ V''''\over  V'' V''})\}.
\eeq
If  $ V(\mu )$  grows as  $\mu ^n$ when  $\mu  \rightarrow  \infty $ ,
then
$ V''''/( V'')^2 \sim  \mu ^{-n}$ , and for our purposes it is
enough to have  $n=p+1 > 1$ , so that in the braces at the $r.h.s$. stands
$\{1+o(1/\mu )\}  (\mu \cdot o(\mu ) \rightarrow  0$ as $\mu
\rightarrow  \infty )$. Then numerator at the $r.h.s$. of (\ref{315}) is
$\sim  1 + o(1/\mu )$, while the second item in square brackets behaves as
$\partial logF(\lambda )/\partial \lambda  \sim  \mu (1+o(1/\mu ))$.
Combining all this, we obtain:

\bea
{\partial \over \partial T_1}\log \ \tau ^{\{ V\}}_1 & = &
res_\mu \left\lbrace  {1+o(1/\mu )\over \mu  - \mu _1} [-
\partial logF(\lambda _1)/\partial \lambda _1 +
\mu (1+o(1/\mu ))]\right\rbrace  = \nn \\
& = & \mu _1 - \partial logF(\lambda _1)/\partial \lambda _1.
\eea
$i.e.$ (\ref{dt1}) is proved for the particular case of  $N =1.$

The proof is literally the same for any  $N$ , we will omit here the
details which can be found in \cite{KMMMZ91b}.
After a simple but cumbersome calculation one gets

\bea
\lefteqn{{\partial \over \partial T_1}\log \ \tau ^{\{ V\}}_N = }\nn \\
& & = res_\mu \left\lbrace {1+o(1/\mu )\over  \prod _{j=1}
(\mu -\mu _j)} \mu ^N\left\lbrace [1+o(1/\mu )] \right. \right. - \nn \\ & & -
\left. \left.
{1\over \mu }[Tr{\partial \over \partial \Lambda _{tr}}\log \ {\rm det} \ F_i(
\lambda _j)]\cdot [1+{\cal O}(1/\mu )]\right\rbrace \right\rbrace = \nn \\
& & =\sum ^N_{j=1}\mu _j -
Tr{\partial \over \partial \Lambda _{tr}}\log \ {\rm det} \ F_i(\lambda _j).
\eea
which completes the proof of eq.(3.13) and thus of the universal
${\cal L}^{\{ V\}}_{-1}$-constraint.

In the particular case of monomial potential $V \equiv V_p =
{X^{p+1}\over p+1}$ (\ref{l-1}) turns into more common form
\cite{FKN1,DVV91a}:

\bea\label{l-1p}
{\cal L}^{\{p\}}_{-1} ={1\over p}\sum _{n\geq 1}
(n+p)T_{n+p} {\partial \over \partial T_n} + \nn \\
+ {1\over 2p}
\sum _{a+b=p \atop{a,b \geq 0}}aT_abT_b
- {\partial \over \partial T_1},
\eea

\bigskip

\section{GKM versus Toda theory and discrete models}

Now, first, without any special
reference to
GKM there exists an explicit relation between KP-like (in Miwa variables),

\beq\label{kp}
\tau _{KP}[T_k] = {\det _{ij}\phi _i(\mu _j)\over \Delta (\mu )},
\eeq
and Toda-like,

\beq\label{tl}
\tau _N[T_{-k},T_k] = \det _{ij} H_{i+N,j+N}[T_{-k},T_k],
\eeq
representations of $\tau $-functions, where

\beq
\Delta (\mu ) =\prod _{i>j}(\mu_ i - \mu _j),
\eeq

\beq
\phi _i(\mu ) = \mu ^{i-1}(1+{\cal O} ({1\over \mu })),
\eeq

\beq\label{miwa2}
T_k = {1\over k} \sum  _i \mu ^{-k}_i\hbox{, }   k>0,
\eeq

\beq\label{toda+}
\partial H_{ij}/\partial T_k = H_{i,j-k}\hbox{, }   j>k>0,
\eeq
and

\beq\label{toda-}
\partial H_{ij}/\partial T_{-k} = H_{i-k,j}\hbox{, }   i>k>0.
\eeq

Relation between (\ref{kp}) and (\ref{tl}) is formulated in terms of the Schur
polynomials, which are defined by:

\beq\label{schur}
{\cal P}[z|T_k] \equiv  \exp \{\sum _{k>0}T_kz^k\} = \sum    z^kP_k[T],
\eeq
$e.g.$  $P_{-n} = 0$  for any $n>0$;  $P_0[T] = 1$;  $P_1[T] = T_1$;
$P_2[T] =
T_2 + {1\over 2}T^2_1$;  $P_3[T] = T_3 + T_2T_1 + {1\over 6}T^3_1$ etc. The
crucial property of Schur polynomials is:

\beq\label{dschur}
\partial P_k/\partial T_n = P_{k-n}
\eeq
(this is just because  $\partial {\cal P}/\partial T_k = z^k{\cal P})$. This
feature allows one to express all the dependence on time-variables of
$H_{ij}[T]$, which satisfies eqs.(\ref{toda+}) and (\ref{toda-}), through
the Schur polynomials:

\beq\label{expan}
H_{ij}[T_{-p},T_p] =
\sum _{^{k\leq i}_{l\geq -j}}P_{i-k}[T_{-p}]H_{kl}P_{l+j}[T_p],
\eeq
where  $H_{kl} \equiv H_{kl}[0,0]$  is already a $T$-{\it independent} matrix.

Let us begin our consideration from the case, when all  $N = T_{-k} =
0$, then look what happens if $N>0$
\footnote{discussion of $N < 0$ can be found in \cite{KMMM92a}},
and introduce $T_{-k}$-variables.

Given the system of basic vectors  $\phi _i(\mu )$  for $i>0$, we put by
definition

\beq\label{oint}
H_{ij}[T_{-k}=0,T_k] = \oint_{z \hookrightarrow  0}
\phi_i(z)z^{-j}{\cal P}[z|T_k]dz\hbox{, }      i>0\hbox{.}
\eeq
The integration contour is around zero and it can be deformed to encircle
infinity and the singularities of ${\cal P}[z]$, if any. If we just substitute
the definition (\ref{schur}) of ${\cal P}[z]$ into (\ref{oint}), we get
(\ref{expan}) with
$P_{k-i}[T_{-m}=0] = \delta _{ki}$ and

\beq
H_{kl} = \oint_{z \hookrightarrow  0}
\phi_k(z)z^ldz.
\eeq
In order to prove the identity between (\ref{kp}) and (\ref{tl})
let us substitute the Miwa transformation (\ref{miwa2}) to (\ref{schur})

$$
{\cal P}[z|T_k] = {\det \ M\over \det (M-Iz)} = \prod  _i {\mu _i\over
(\mu _i-z)} =
\left[ \prod  _i \mu _i\right]  \sum  _k
{(-)^k\over (z-\mu _k)}{\Delta _k(\mu )\over \Delta (\mu )},
$$
where $\Delta _k(\mu ) \equiv \prod _{\atop {i>j;}{i,j\neq k}}(\mu
_i-\mu _j)$.
Now the integral (\ref{oint}) picks up contributions only from the poles of
${\cal P}[z|T_k]$
at the points $\mu _k$:

$$
H_{ij}[T_{-k}=0,T_k] = \oint_{z \hookrightarrow 0}
\phi_i(z)z^{-j}{\cal P}[z|T_k]dz = {
\prod_i \mu _i\over \Delta (\mu )} \sum  _k (-)^k \phi _i(\mu _k){\Delta
_k(\mu )\over
\mu ^j_k}.
$$
The sum at the $r.h.s.$ has a form of a matrix product and we conclude that

$$
\det \ H_{ij} = \det \ \phi _i(\mu _k)\cdot \prod  _k \left[ {
\prod _i \mu _i\over \Delta (\mu )} (-)^k \Delta _k(\mu )\right] \cdot \det
{1\over \mu ^j_k}.
$$
The last determinant at the $r.h.s.$ is equal to  $\Delta (1/\mu ) \sim
\Delta (\mu )\cdot \left[ \prod  _k \mu ^N_k\right] ^{-1}$. Note also
that  $\displaystyle{\prod  _k \left[ {\Delta _k(\mu )\over \Delta (\mu )}
\right]  =
\Delta (\mu )^{-2}}$, and taking all this together, we see that there
exists the equality

\beq
\det \ H_{ij} = {\det \ \phi _i(\mu _j)\over \Delta (\mu )},
\eeq
as required.

Proceed now to introducing of zero- and negative-time variables. The zero-time
$n$  arises just as the simultaneous shifts of indices $i$ and $j$ of $H_{ij}:
H_{ij} \rightarrow  H_{i+N,j+N}$ . We can write:

\beq
H_{i+N,j+N}[0,T_k] = \oint_{z \hookrightarrow 0}
\phi^{\{N\}}_i(z)z^{-j}{\cal P}[z]dz
\eeq
with

\beq
\phi ^{\{N\}}_i(z) = z^{-N}\phi _{i+N}(z)
\eeq
This exhausts the problem of restoring the $N$-dependence for positive integer
values of $N$.

As for negative-times, as soon as $H_{kl}$ is defined, they are introduced with
the help of (\ref{expan}) and

\beq\label{full}
\new
\begin{array}{c}
H_{i+N,j+N}[T_{-k},T_k] \equiv \sum _{k\leq
i}P_{i-k}[T_{-p}]H_{k+N,j+N}[0,T_l]
=\\
= \oint_{z \hookrightarrow 0}
\phi^{\{T_{-k},N\}}_i(z)z^{-j}{\cal P}[{1\over z}|T_{-k}]{\cal P}[z|T_k]dz,
\end{array}
\eeq
with

\beq\label{phi-}
\new
\begin{array}{c}
\phi ^{\{T_{-k},N\}}_i(z) \equiv
\left\lbrace {\cal P}[{1\over z}|T_{-k}]\right\rbrace ^{-1} \sum
P_{i-k}[T_{-l}]\phi ^{[N]}_k = \\
= z^{-N}\exp \left\lbrace -\sum _{k>0}T_{-k}z^{-k}\right\rbrace  \sum
P_k[T_{-l}]\phi _{i+N-k}(z).
\end{array}
\eeq
The role of the exponential prefactor in (\ref{phi-}) is to guarantee the
proper
asymptotic behaviour

\beq
\phi ^{\{T_{-k},N\}}_i(z) = z^{i-1}\{1+{\cal O} ({1\over z})\}.
\eeq

The important reduction from Toda lattice is Toda chain
(see, for example, \cite{UT84}). It can be easily
written both in terms of element $G$ in the fermionic language ($[G,J_k +
J_{-k}]=0$ -- see also the comments above,
and in determinant form. Latter one merely implies the
symmetry property:

\beq\label{chain}
[H\hbox{ , } \Lambda +\Lambda ^{-1}]=0,
\eeq
where $\Lambda $ is shift matrix  $\Lambda _{ij}\equiv \delta _{i,j-1}$. This
condition leads to $\tau $-function of Toda chain hierarchy (proper rescaled by
exponential of bilinear form of times) which depends only on the sum of
positive and negative times  $t_k = {1\over 2}(T_k+T_{-k})$,
but not on their difference (one can
consider this as defining property of Toda chain hierarchy). Let us remark that
one possible solution to constraint (\ref{chain}) is matrix
$H_{i,j}={\cal H}_{i-j}$.
We can combine both reductions to
reproduce forced Toda chain hierarchy. In this case one can easily transform
$H_{i-j}$
to matrix $\tilde {\cal H}_{i+j}$ by permutations of columns what does not
effect to the determinant. This matrix just corresponds to one-matrix model
case \cite{GMMMO,KMMOZ}.
Thus, we consider again the
determinant of size $N\times N$, which now can be represent in the form:

\beq
\tau _N = \det_{N\times N}
\partial ^{i+j}H
\eeq
where $\partial \equiv \partial /\partial t_1$,
$\partial H/\partial t_k=\partial ^kH$.
Like the Toda lattice case, the forced Toda chain is unambiguously
continuable to negative values of zero-time.

Now, we can easily introduce zero- and negative-time
variables into GKM in such a way that its partition function becomes a
$\tau $-function of the Toda lattice hierarchy.
The relevant
set of functions $\{\phi _i(\mu )\}$ -- the point in Grassmannian for
GKM -- is given by
the following integral formula:

\beq
\new
\begin{array}{c}
\phi ^{\{V\}}_i(\mu ) = e^{V(\mu )-\mu V'(\mu )}\sqrt{V''(\mu )}\int
dx\ x^{i-1}e^{-V(x)+xV'(\mu )} \equiv \\
\equiv  s(\mu )\int   dx\ x^{i-1}e^{-V(x)+xV'(\mu )} \equiv  \left< x^{i-1}
\right> _\mu
\end{array}
\eeq
Dependence of  $N$  and $T_{-k}$ is now introduced by the rule
\footnote{Let us point out that the exponential of negative powers in
normalization does not essentially effect to the KP $\tau$-function as it
reduces to trivial exponential of bilinear form of times in front of
$\tau$-function and corresponds to the freedom in its definition.
Indeed, $\tau \sim \det \left\{ \exp
[\sum _ka_kz^{-k}_j]\phi _i(z_j)\right\}\sim \prod _l \exp [\sum a_kz^{-k}_l]
\det \phi _i(z_j) \sim \exp [\sum ka_kT_k] \det \phi _i(z_j)$.
}:

\beq\label{psigkm}
\new
\begin{array}{c}
\phi _i ^{\{V,N,T_{-k}\}}(\mu ) \equiv  \left<
x^{i-1}\left[ {x\over \mu }\right] ^N\exp \left\lbrace
\sum _{l>0}T_{-l}(x^{-l
}-\mu ^{-l})\right\rbrace  \right> _\mu  = \\
= {\sqrt{V''(\mu )} e^{V(\mu )-\mu V'(\mu )}\over \mu ^N}\int
dx\ x^{N+i-1}e^{-V(x)+xV'(\mu )}\exp \left\lbrace \sum
_{l>0}T_{-l}(x^{-l}-\mu ^{
-l})\right\rbrace  = \\
= e^{\hat V(\mu )-\mu V'(\mu )}\sqrt{V''(\mu )}\int   dx\ x^{i-1}e^{-\hat V(x)+
xV'(\mu )},
\end{array}
\eeq
where

\beq
\hat V(X) \equiv  V(X) - N\log X - \sum _{k>0}t_{-k}X^{-k}
\eeq
with

The original potential $V$ it can be identified with $\hat V_+$.
{}From (\ref{psigkm}) we immediately conclude
that the partition function of
GKM, involving
zero- and negative-times (and automatically being a Toda lattice
$\tau $-function), is just

\beq
\hat Z_{\{\hat V\}}[M] = e^{Tr\hat V(M)-TrM\hat V'_+(M)}
{\int DX\  e^{-Tr\hat V(X)+Tr\hat V'_+(M)X}\over
\int dX\ e^{-Tr\hat U_{+,2}(X,M)}}
\eeq

Since we devote this section to discussion of Toda lattice hierarchies in the
context of matrix models, we can not avoid touching the main conclusion of
\cite{GMMMO} that all the {\it discrete} matrix models do correspond to
particular
cases of Toda hierarchies. In the simplest case of Hermitean one-matrix model
one
gets a Toda chain, other multi-matrix models correspond to other reductions of
the Toda lattice hierarchy. Moreover, all discrete matrix models fall into the
class of {\it forced} hierarchies \cite{KMMOZ}.

The idea is to perform a Miwa transformation of times $T_k$ with $T_{-k}$
fixed, so that  $H_{ij}$'s become averages of polynomial functions of {\it X}.
Then  $\det \ H_{ij}$ may be transformed with the help of orthogonal
polynomials technique.

The main result of all these calculations is that partition functions arise in
the form (\ref{kp}), with  $\phi _i(z)$  being proportional to orthogonal
polynomials.

Indeed, using

\beq\label{miwa3}
T_k = {1\over k} Tr \Lambda ^{-k} = {1\over k}\sum ^{\tilde N}_{i=1}
\lambda ^{-k}_i +
\tilde t_k
\eeq
($k > 0$)
\footnote{
and/or
$$
T_{-k} \equiv  \bar T_k = {1\over k}TrM^{-k} = {1\over k}\sum ^{\tilde
N}_{i=1}
\mu ^{-k}_i + \tilde T_{-k}
$$
}
(note that  $\tilde N$ -- the size of matrix  $\Lambda $  has nothing to do
with  $N$ --
the size of matrices  $M$, being integrated over in (\ref{z2}) \ )
the partition functions of discrete models  $\{\tau _N\}$
acquire the form of KP $\tau $-function (\ref{kp}). $E.g.$, for the
Hermitean one-matrix model one has

\beq\label{vyvod}
\new
\begin{array}{c}
\tau _N(t) = (N!)^{-1}\int   \prod  _i dm_i \Delta ^2(m) \exp
\{-\sum _{i,k}t_{_k}m^{k}_i \} = \\
= (N!)^{-1}\int   \prod  _i dm_i \Delta ^2(m) e^{-\tilde V(m_i)}
\prod _{i,a}(1 -
{m_i\over \lambda _a}) = \\
=(N!)^{-1}\prod  _a \lambda ^{-N}_a\int   \prod  _i dm_i e^{-\tilde V(m_i)}
\Delta (m)
{\Delta (m,\lambda )\over \Delta (\lambda )} = \\
= (N!)^{-1}\prod  _a \lambda ^{-N}_a \Delta ^{-1}(\lambda )\int   \prod  _i
dm_i
e^{-\tilde V(m_i)} \times \\
\times  \det _{N\times N} \tilde P^{(1)}_{i-1}(m_j) \det _{(N+\tilde N)\times
(N+\tilde N)}
\left[ \begin{array}{rcl}
\tilde P^{(l)}_{i-1}(m_j) & \vdots & \tilde P^{(l)}_{N+b-l} (m_j)\\
\ldots & \vdots & \ldots\\
\tilde P^{(l)}_{i-1}(\lambda _a) & \vdots & \tilde P^{(l)}_{N+b-l}
(\lambda _a)
\end{array}
\right]\hbox{ , }
\end{array}
\eeq
where  $i,j = 1,\ldots,N$;  $a,b = 1,\ldots,\tilde N$;
and  $\{\tilde P_i(m)\}$  are corresponding
orthogonal polynomials with respect to deformed measure  $e^{-\tilde V}
dm:$

\beq\label{orth}
<\tilde P_i,\tilde P_j> = \int
\tilde P_i(m)\tilde P_j(m)e^{-\tilde V(m)}dm =
\delta _{ij}e^{\tilde \varphi _i(s)}\hbox{ , }
\eeq
so that

$$
\tilde P_i(m) = m^i + O(m^{i-1}).
$$
Computing determinants in (\ref{vyvod}) and using orthogonality condition
(\ref{orth}) one obtains

\beq\label{rep}
\new
\begin{array}{c}
\tau _N[\lambda |\tilde t] = \prod  _a \lambda ^{-N}_a \Delta ^{-1}(\lambda )
\det _{\tilde N\times \tilde N}
\tilde P_{N+a-1}(\lambda _b) \prod  _i e^{\tilde \varphi _i(\tilde t)} = \\
= \left[ \prod  _i e^{\tilde \varphi _i(\tilde t)}\right]
{\det _{(ab)}\phi ^{(N)}_a(\lambda _b)\over \Delta (\lambda )} =
\tau _N[\tilde t]
\times {\det _{(ab)}\phi ^{(N)}_a(\lambda _b)\over \Delta (\lambda )}\hbox{ ,}
\end{array}
\eeq
$i.e.$ the $\tau $-function of the discrete Hermitean one-matrix model acquires
the form of eq.(\ref{kp}) with

\beq
\phi ^{(N)}_a(\lambda ) = \lambda ^{-N} \tilde P_{N+a-1}(\lambda )
\eeq
(\ref{rep}) is natural representation for {\it all} discrete
matrix models.

Now, in order to be representable as GKM the above formulas still need to
arise in a somewhat specific form. Namely, components of the vector
$\{\phi _i(\mu )\}$ should possess a representation as ``averages",

$$
\phi _i(\mu ) = \left< x^{i-1} \right> _\mu
$$
Since in the study of discrete matrix models  $\phi _i(\mu )$  arise as
the orthogonal
polynomials  $P_{n+i}(\mu )$, what is necessary is a kind of integral
representation of these polynomials, with $i$-dependence coming only from the
$x^{i-1}$--factor in the integrand. It is an interesting problem to find out
such kind of representation for various discrete models, but it is easily
available only whenever orthogonal polynomials are associated with the Gaussian
measure: the relevant Hermit polynomials are known to possess integral
representation, which is exactly of the form which we need.

The final statement we have got here is that Hermitean one-matrix model with
the matrix
size $N$ is equivalent to GKM with $\hat V(X) = X^2/2 - N\log X$.

Indeed, we have proven the remarkable
explicit identity between Gaussian matrix
integrals,

\beq\label{id}
{\int DM_{N\times N}\det(I-M/\Lambda ) e^{-TrM^2/2}\over
\int DM_{N\times N} e^{-TrM^2/2}}  =  {\int
DX_{\tilde N\times \tilde N}\det (I-iX/\Lambda )^N e^{-TrX^2/2}\over
\int DX_{\tilde N\times \tilde N}e^{-TrX^2/2}}
\eeq
Note that the size of matrix in the $l.h.s.$ is $N\times N$ and in the $r.h.s.$
is $\tilde N\times \tilde N$, and these parameters are absolutely {\it
independent}.
This identity
is indeed true for any  $N$  and  $\tilde N$,
as follows from the proof given above
and integral representation of Hermit polynomials
$\phi _j(\mu ) = He_j(i\mu )$.

\section{Double-scaling limit}

Now, we are going directly to the discussion of the connection between
discrete and continuum theories. Indeed, it turns out \cite{MMMM91} that
this connection at the language of free scalar fields is nothing but a
change of spectral parameter

\beq\label{spec}
u^2 = 1 + az
\eeq
Then the continuous Virasoro constraints (\ref{dvir})
which are modes of the stress tensor
\beq\label{b3}
{\cal T}(z)= {1\over 2}{:}\partial\Phi^{2}(z){:} - {1\over16z^2}
=\sum{{\cal L}_n\over z^{n+2}}.
\eeq
can be deduced \cite{MMMM91} from analogous constraints in Hermitian one-\mm
by taking the continuum limit.
The procedure is as follows.

The partition function of Hermitian one-\mm  (\ref{mamo})
satisfies the discrete Virasoro constraints
(\ref{dvir}).
In order to obtain the
continuum constraints (\ref{vir}) one has to consider a reduction of  model
(\ref{z2}) to the pure even potential $t_{2k+1}=0$.

Let us denote by the $\tau_N^{\rm red}$ the partition function of the
reduced \mm
\beq	\label{a1}
\tau^{\rm red}_{N}\{t_{2k}\}=\int{\cal D}M\exp{\rm Tr}\sum_{k=0}t_{2k}
M^{2k}
\eeq
and consider the following change of the time variables
\beq\label{a5}
g_m=\sum_{n\geq m}{(-)^{n-m}\Gamma\left(n+{3\over2}\right)
a^{-n-{1\over2}}\over(n-m)!\Gamma\left(m+{1\over2}\right)}T_{2n+1},
\eeq
($g_m \equiv mt_{2m}$ and this expression can be used also for the zero
discrete time $g_0 \equiv N$ that plays the role of the dimension of matrices
in the one-matrix model),
deduced from the following prescription. Take the free
 scalar field with periodic boundary conditions for $p=2$)
\beq\label{a8}
\partial\varphi(u)=\sum_{k\geq0}g_ku^{2k-1}+
\sum_{k\geq1}{\partial\over\partial t_{2k}}u^{-2k-1},
\eeq
and analogous scalar field with antiperiodic boundary conditions:
\beq\label{a9}
 \partial\Phi(z)=\sum_{k\geq0}\left(\left(k+{1\over2}\right)T_{2k+1}
z^{k-{1\over2}}+
{\partial\over\partial \tilde T_{2k+1}}z^{-k-{3\over2}}\right).
\eeq
Then the equation
\beq\label{a10}
{1\over\tilde\tau}\partial\Phi(z)\tilde\tau =
a {1\over\tau^{\rm red}}\partial\varphi(u)\tau^{\rm red},\quad
u^2=1+az
\eeq
generates the correct transformation rules
(\ref{a5}), (\ref{a6}) and gives rise to the expression
for $A_{nm}$ (\ref{a12}).
Taking  the square of the both sides of the identity (\ref{a10}),
\bea\label{a100}
&{1\over\tilde\tau}{\cal T}(z)\tilde\tau={1\over \tau^{\rm red}}T(u)\tau^{\rm
red},
\eea
one can obtain after simple  calculations that the
relation (\ref{a15}) is valid.

Derivatives with respect to  $t_{2k}$ transform as
\beq\label{a6}
{\partial\over\partial t_{2k}}=\sum_{n=0
}^{k-1}{\Gamma\left(k+{1\over2}\right)
a^{n+{1\over2}}\over(k-n-1)!\Gamma\left(n+{3\over2}\right)}{\partial
\over\partial \tilde T_{2n+1}},
\eeq
where the auxiliary continuum times $\tilde T_{2n+1}$ are connected with
``true'' Kazakov continuum times $T_{2n+1}$ via
\beq\label{a7}
T_{2k+1}=\tilde T_{2k+1}+a{k\over k+1/2}\tilde T_{2(k-1)+1},
\eeq
and coincide with $T_{2n+1}$ in the double-scaling limit when $a\to 0$.

Let us rescale the partition function of the reduced one-\mm by
exponent of quadratic form of the auxiliary times $\tilde T_{2n+1}$
\beq\label{a11}
\tilde\tau=\exp\left(-{1\over2}\sum_{m,n\geq0}A_{mn}\tilde T_{2m+1}
\tilde T_{2n+1} \right)\tau^{\rm red}_N
\eeq
with
\beq\label{a12}
A_{nm}={\Gamma\left(n+{3\over2}\right)\Gamma\left(m+{3\over2}\right)\over
2\Gamma^2\left({1\over2}\right)}
{(-)^{n+m}a^{-n-m-1}\over n!m!(n+m+1)(n+m+2)}.
\eeq
Then a direct though tedious calculation \cite{MMMM91} demonstrates that
the relation
\beq\label{a15}
{\tilde{\cal L}_n\tilde\tau\over\tilde\tau}
=a^{-n}\sum_{p=0}^{n+1}C^p_{n+1}(-1)^{n+1-p}
{L_{2p}^{\rm red}\tau^{\rm red}\over \tau^{\rm red}},
\eeq
is valid, where
\bea
L_{2n}^{\rm red} \equiv \sum_{k=0}kt_{2k}{\partial\over\partial t_{2(k+n)}}+
\sum_{0\leq k\leq n}{\partial ^2\over \partial t_{2k}\partial t_{2(n-k)}}
\eea
and
\bea\label{b6}
\tilde{\cal L}_{-1}=&\sum_{k\geq1}\left(k+{1\over2}\right)T_{2k+1}
{\partial\over\partial \tilde T_{2(k-1)+1}}+
{T_1^2\over16G},\nn\\
\tilde{\cal L}_{0}=&\sum_{k\geq0}\left(k+{1\over2}\right)T_{2k+1}
{\partial\over\partial \tilde T_{2k+1}},\nn\\
\tilde{\cal L}_n=&\sum_{k\geq0}\left(k+{1\over2}\right)T_{2k+1}
{\partial\over\partial \tilde T_{2(k+n)+1}} \nn\\
&+\sum_{0\leq k \leq n-1}{\partial\over\partial \tilde T_{2k+1}}
{\partial\over\partial \tilde T_{2(n-k-1)+1}} -{(-)^n\over16a^n},\ \ \
n\geq1.
\eea
Here $C^p_n =\frac{n!}{p!(n-p)!}$ are binomial coefficients.

These Virasoro generators differ from (\ref{vir})
by terms which are singular in the limit $a\rightarrow 0$.
At the same time $L_{2p}^{\rm red}\tau^{\rm red}$ at the r.h.s. of (\ref{a15})
do
not need to vanish, since
\bea\label{15aa}
0 = L_{2p}\tau\left\vert_{t_{2k+1}=0} =
L_{2p}^{\rm red}\tau^{\rm red} +
\sum_i {\partial^2\tau\over\partial t_{2i+1}\partial t_{2(n-i-1)+1}}
\right\vert_{t_{2k+1}=0}.
\eea
It was shown in \cite{MMMM91} that these two origins of difference between
(\ref{vir}) and (\ref{b6}) actually cancel each other, provided eq.(\ref{a15})
is rewritten in terms of the square root $\sqrt{\tilde\tau}$ rather than
$\tilde\tau$ itself:
\beq\label{a16}
{{\cal L}^{\rm cont}_n\sqrt{\tilde\tau}\over\sqrt{\tilde\tau}}
=a^{-n}\sum_{p=0}^{n+1}C^p_{n+1}(-1)^{n+1-p}\left. {L_{2p}\tau\over\tau}
\right\vert_{t_{2k+1}=0}\left( 1+  O(a) \right).
\eeq
The proof of this cancelation, as given in  \cite{MMMM91}, is not too much
simple and makes use of integrable equations for $\tau$.

Now we will use the demonstrated above fact, that the discrete Hermitean
one-matrix
model is equivalent to GKM with  $\hat V(X) = X^2/2 - N\log X$, and
also that its
double-scaling continuum limit is described by GKM with  $V(X) = X^3/3$. Thus,
we should conclude that

\beq\label{ds}
\lim_{d.s.\hbox{ }N\rightarrow \infty }Z_{\{\hat V\}} = Z^2_{\{V\}}.
\eeq
This relation should certainly be understandable just in terms of GKM itself.

Let us recall that double-scaling continuum limit for the model
of interest implies that only even times
$\displaystyle{t_{2k} = {1\over 2k} Tr
{1\over \Lambda ^{2k}}}$  should remain non-zero, while all odd times
$t_{2k+1} = 0$.
This obviously implies that the matrix $M$ should be of the block form:

\beq\label{block}
\Lambda = \left(
\begin{array}{cc}
{\cal M} & 0\\0 & -{\cal M}
\end{array}\right)
\eeq
and, therefore, the matrix integration variable is also naturally decomposed
into block form:

\beq
X = \left(
\begin{array}{cc}
{\cal X} & {\cal Z}\\{\cal Z} & {\cal Y}
\end{array}
\right) .
\eeq
Then

\beq
\new
\begin{array}{c}
Z_{\{\hat V=X^2/2-N\log X\}} =\\
= \int D{\cal X}D{\cal Y}D^2{\cal Z}\ \det ({\cal X}{\cal Y}-\bar {\cal Z}
{1\over {\cal Y}}{\cal Z}{\cal Y})^N
e^{-Tr\{|{\cal Z}|^2+{\cal X}^2/2+{\cal Y}^2/2-{\cal M}{\cal X}+
{\cal M}{\cal Y}\}}.
\end{array}
\eeq
To take the limit $N\rightarrow \infty $, one should assume certain scaling
behaviour of ${\cal X}$, ${\cal Y}$ and ${\cal Z}$. Moreover,
the notion of {\it double}-scaling
limit implies a specific {\it fine tuning} of this scaling behaviour. So we
shall take

\beq
\new
\begin{array}{c}
{\cal X} = \alpha (i\beta I + x), \\
{\cal Y} = \alpha (-i\beta I + y),      \\
{\cal Z} = \alpha \zeta ,
\end{array}
\eeq
with some large real $\alpha $ and $\beta $.

As for behaviour of the matrix of Miwa's parameters, it should be
dictated by the change of the spectral parameter (\ref{spec}), which
dictates the right Kazakov change of variables. So, we will take the
anzatz
\beq\label{anz}
{\cal M}^2 = A + Bm
\eeq
which under particular scaling behaviour $A/B \stackreb{N \to
\infty}{\to} \infty$ could be linearized giving rise to
\beq\label{anzlin}
{\cal M} = \alpha ^{-1}(i\gamma I + m)
\eeq
with
\beq
A^{1/2} = {\gamma \over \alpha}
$$
$$
{B \over 2A^{1/2}} = {1\over \alpha}
\eeq

If expressed through
these variables, the action becomes:

\beq
\new
\begin{array}{c}
Tr\{|{\cal Z}|^2 + {\cal X}^2/2 + {\cal Y}^2/2 - {\cal M}{\cal X} +
{\cal M}{\cal Y} - N\log ({\cal X}{\cal Y} - \bar {\cal Z}{1\over
{\cal Y}}{\cal Z}{\cal Y})\} = \\
= {\gamma ^2\over 2}Tr\{(i\beta I + x)^2 + {\gamma ^2\over 2}Tr(i\beta I -y)^2
+ \gamma ^2|\zeta |^2\} - Tr(i\alpha I + m)(2i\beta I + x - y) - \\
- NTr \log \ \beta ^2\gamma ^2\{1 - i {x-y\over \beta } +
{xy\over \beta ^2} - {|\zeta |^2\over \beta ^2}(1 + o(1/\beta ))\} =
\end{array}
\eeq
$$
= [2\alpha \beta  - \beta ^2\gamma ^2 - 2N\ \log \ \beta \gamma ] Tr\ I -
2i\beta \ Tr\ m +
\eqno{(A)}
$$
$$
+ i(\beta \gamma ^2 - \alpha  + {N\over \beta })(Tr\ x - Tr\ y) +
{1\over 2}(\gamma ^2 - {N\over \beta ^2})(Tr\ x^2 + Tr\ y^2) +
\eqno{(B)}
$$
$$
+ (\gamma ^2 - {N\over \beta ^2}) Tr |\zeta |^2 -
\eqno{(C)}
$$
$$
- Tr mx + Tr\ my + {iN\over 3\beta ^2}Tr(x^3 - y^3) +
\eqno{(D)}
$$
$$
+ {\cal O}(N/\beta ^4) + {\cal O}(|\zeta |^2 {N\over \beta ^3}).
\eqno{(E)}
$$
We want to adjust the scaling behaviour of $\alpha $, $\beta $ and $\gamma $ in
such a way that only the terms in the line $(D)$ survive. This goal is achieved
in several steps.

The line ({\it A}) describes normalization of functional integral, it does not
contain $x$ and $y$. Thus, it is not of interest for us at the moment.

Two terms in the line $(B)$ are eliminated by adjustment of $\alpha $ and
$\gamma $:

\beq
\gamma ^2 = {N\over \beta ^2}\hbox{ , }  \alpha  = {2N\over \beta }\hbox{ .}
\eeq
As we shall see soon,  $\gamma ^2 = N/\beta ^2$ is large in the limit of
$n\rightarrow \infty $ . Thus, the term $(C)$ implies that the fluctuations of
$\zeta $-field are severely suppressed, and this is what makes the terms of the
second type in the line $(E)$ negligible. More general, this is the reason for
the integral  $Z_{\{\hat V\}}$
to split into a product of two independent integrals leading to the square
of partition function in the limit $n\rightarrow \infty $ (this splitting is
evident as, if ${\cal Z}$ can be neglected, the only mixing term
$\displaystyle{\log \det
\left(
\begin{array}{cc}
{\cal X} & {\cal Z}\\{\cal Z} & {\cal Y}
\end{array}
\right)} $  turns into  $\log {\cal X}{\cal Y} = \log {\cal X} +
\log {\cal Y}$).

Thus, we remain with a single free parameter $\beta $ which can be adjusted so
that

\beq
{\beta ^3\over N}\rightarrow const\ \ \
\hbox{   as }\ \ \ N\rightarrow \infty \ \ \
(i.e\hbox{. }  \beta  \sim  N^{1/3}  ),
\eeq
making the terms in the last line $(E)$ vanishing and the third term in the
line $(D)$ finite.

Let us finish the discussion of double-scaling limit, making some
remarks on ref. \cite{ACKM}. It is claimed there that in order
to get the Kontsevich model from a discrete Hermitean one, one should
not necessarily care of the reduction to even times (\ref{block}) and
the particular Kazakov change of variables, inspired by (\ref{spec}),
(\ref{anz}); instead it is just enough to take (\ref{anzlin}) and make
the rescalings similiar to what we done above. However, in such case it
is not clear what should be instead of Kazakov change of variables, and
what is indeed the integrable hierarchy we get in continuum limit. As
for the first question, one might hope that the related change of
variables in in the class of ''allowed redefinitions" of Kazakov times,
similiar to those we used in  (\ref{a7}), though the second question,
certainly deserves further investigation.

\section{GKM as a solution to topological $(p,1)$ models}

For various choices of the potential  $V(X)$  the model (\ref{gkm}) formally
reproduces various
$(p,q)$-series:  the potential $V(X) = {X^{p+1}\over p+1}$  can be associated
with the entire set of $(p,q)$-minimal string models with all possible
$q$'s.
In order to specify $q$ one needs to make a special choice of $T$-variables:
all  $T_k= 0$, except for  $T_1$ and  $T_{p+q}$ (the symmetry between $p$ and
$q$ is implicit in this formulation).

Let us briefly discuss two simple examples. First, we will fix $p=2$,
$i.e.$ the case of the {\it KdV} reduction to the KP hierarchy. The
second number $q$ should be coprime to $p$, thus we have here $q =
2m-1$.

Now the string equation is of the form
\bea
{1\over \tau _{KdV}}{\cal L} _{-1} \tau _{KdV} = {1\over 2}
\sum _{{k>1 }\atop {k\ odd}}
kT_k {\partial \over \partial T_{k-2}}\log \tau _{KdV} +
{T_1^2 \over 4} = 0
\eea
or taking the ${\partial \over \partial T_1}$-derivative, one gets

\beq\label{stre}
\sum _{{k>1 }\atop {k\ odd}}
kT_k {\partial ^2\over \partial T_{k-2} \partial T_1}\log \tau _{KdV} +
T_1 = 0
\eeq
or using the formula for the Gelfand-Dikii polynomials

\beq
{\partial ^2\over \partial T_{k-2} \partial T_1}\log \tau _{KdV} =
\left[ L^{2m-1} \right]_{-1} \equiv {\cal R} _m [u]
\eeq
we have

\beq\label{gd}
\sum _{m \geq 0} (2m+1)T_{2m+1} {\cal R} _m [u] = 0
\eeq

Now, we should use the ''axiomatics" of \cite{FKN1} how to extract the
concrete $(2,2m-1)$ solutions from (\ref{stre}), (\ref{gd}). The very
simple example is $m=1$:
where
\beq
3T_3 {\partial ^2 \over \partial T_1^2} \log \tau _{KdV} + T_1 = 0
\eeq
using that
$$
u \sim  {\partial ^2 \over \partial T_1^2} \log \tau _{KdV}
$$
the solution to the KdV-equation is

\beq
u \sim {T_1\over T_3}
\eeq
or fixing $T_3$, one gets

\beq\label{c=-2}
F = \log \tau \sim T_1^3
\eeq
which is a well known fact from the $c=-2$ theory coupled to gravity
where
$$
\langle P^3 \rangle = 1
$$
with $P$ being the puncture operator $P = c \bar c e^{\phi}$. This is
the case of {\it topological} gravity.

Less simple example is with $m=2$,
where (\ref{gd}) and the fact that
$$
{\cal R}_2 \sim u^2 + u''
$$
gives the Painleve equation

\beq\label{painleve}
u^2 + u'' = T_1
\eeq

This is the case of {\it pure} gravity where the solution is actually much more
complicated than in the previous case.

{}From this point of view, the presense of {\it all} $(p,q)$ solutions in
GKM  is a rather formal consideration. For the potential  $V(X) =
{X^{p+1}\over p+1}$  the partition function  $Z[V|T_k] = \tau _V[T_k] \equiv
\tau _p[T_k]$  satisfies the string equation which looks like

\beq\label{4}
\sum ^{p-1}_{k=1}k(p-k)T_kT_{p-k} + \sum ^\infty _{k=1}(p+k)(T_{p+k} -
{p\over p+1}\delta _{k,1}) {\partial \over \partial T_k} \log \ \tau _p[T] = 0
\eeq
$i.e$. $\tau $-function is defined with all Miwa times (\ref{miwa}) around zero
values
(in $1/M$ decomposition like in original Kontsevich model) with the only
exception - $T_{p+1}$ is shifted what corresponds obviously to $(p,1)$
model.
Thus, we see that the matrix integral gives an explicit solution to
$(p,1)$ string models which must be nothing but particular topological matter
coupled to topological gravity.

Of course, we still have an opportunity for analytic continuation in string
equation, using the definition of Miwa's times (\ref{miwa}). We have to
satisfy the following conditions:
\beq\label{5}
T_1 = x
$$
$$
T_2 = 0
$$
$$
...
$$
$$
T_{p+1} - {p\over p+1} = 0
$$
$$
T_{p+q} = t_{p+q} = fixed
$$
$$
T_{p+q+1} = 0
$$
$$
...
\eeq
which is a system of equations on the Miwa parameters  $\{\mu _i\}$,  $i =
1,...,N$. So, to do this analytic continuation one has to decompose the whole
set
\beq\label{6}
\{\mu _i\} = \{\xi _a\} \oplus  \{\mu '_s\}
$$
$$
T_k = {1\over k} TrM^{-k} = {1\over k} \sum ^N_{j=1}\mu ^{-k}_j = {1\over k}
\sum    \xi ^{-k}_a + {1\over k} \sum ^{N'}_{j=1}{\mu '}_j^{-k} \equiv
T^{(cl)}_k + T'_k
\eeq
into ``classical" and ``quantum" parts respectively. In principle it is clear
that we have now to solve the equations
\beq\label{7}
T^{(cl)}_k = {1\over k} \sum    \xi ^{-k}_a = t_{p+q}\delta _{k,p+q} -
{p\over p+1}\delta _{k,p+1}
\eeq
and this can be done adjusting a certain block form of the matrix  $M$
\cite{KMMMZ91b,Mar92}. However, in such a way we can only vanish several first
times, the rest ones can be vanished only adjusting correct behaviour in the
limit $N \rightarrow  \infty $. The most elegant way to do this
\footnote{due to A.Zabrodin}
is to use the formula
\beq\label{8}
\exp  (- \sum ^\infty _{k=1}\lambda ^kT^{(cl)}_k) =
\lim _{K \rightarrow \infty }
(1 - {1\over K}\sum ^\infty _{k=1}\lambda ^kT^{(cl)}_k)^K = \prod  _a (1 -
{\lambda \over \xi _a})
\eeq
and then the solution to (\ref{7}) will be given by  $K$  sets of roots of the
equation
\beq\label{9}
\sum ^\infty _{k=1}\lambda ^kT^{(cl)}_k - K = t_{p+q}\lambda ^{p+q} -
{p\over p+1}\lambda ^{p+1} - K = 0
\eeq
Obviously, the eigenvalues  $\xi _a$ will now depend on the size of the matrix
$N = (p+q)K + N'$  through explicit  $K$-dependence  $(\xi _a \sim
K^{1/(p+q)})$  and we lose one of the main features of $(p,1)$
theories mentioned above -- trivial dependence of the size of the matrix. Now
we can consider only matrices of {\it infinite} size and deal only with the
infinite determinant formulas.

That is why this way to get higher critical points is a formal one.
Below we will present an alternative way of thinking
\cite{KMMM92b,KM1,KM2}, connected with
so-called $p$-times. Indeed, there
exists {\it a
priori} another integrable structure in the model (\ref{gkm}), connected with
time
variables, related to the non-trivial coefficients of the potential  V. As a
results, the cases of monomial potential  $V_p(X) = {X^{p+1}\over p+1}$  and
arbitrary polynomial of the same degree  $(p+1)$  are closely connected with
each other.

In order to demonstrate this, first, we return to
the derivatives of  $Z_{GKM}$ with respect to the time-variables  $T_k$. Such
derivatives define nonperturbative correlators in string models and are of
their own interest for the theory of GKM. The derivatives with respect to
$T_k$ with  $k \geq  p+1$  (responsible for the correlators of irrelevant
operators) are not very easy to evaluate, things are simpler for  $T_k$ with
$1 \leq  k \leq  p$, where
using the obvious notation of average so that  $Z_{GKM} =
\left< 1\right> $, we have
\beq\label{lgder}
\left.{\partial Z_{GKM}\over \partial T_k}\right| _V = \left< TrM^k -
TrX^k\right> , \ \ \  1 \leq  k \leq  p
\eeq
It is implied that the derivative in the $l.h.s.$ is taken
preserving the form of the potential  $V = \sum ^{p+1} {v_k \over k}X^k$.

The $r.h.s$. of (\ref{lgder}) can be also represented as
\beq\label{der}
\left.{\partial Z_{GKM}\over \partial T_k}\right| _V =
\left<Tr{\partial V(M)\over \partial v_k} -
Tr{\partial V(X)\over \partial v_k}\right> \hbox{, }   1 \leq  k \leq
p\
\eeq
which looks similar but actually {\it is different} from  $-
{\partial \over \partial v_k}Z_{GKM}$, as it would be if (\ref{der}) does
contain some corrections.
The problem is that  ${\partial \over \partial v_k}Z_{GKM}$ gets contributions
not only from differentiating $V(X) - V(M)$ in exponentials in
(\ref{gkm}) but
also from the term  $V'(M)(X-M) \equiv W(M)(X-M)$ as well as from the
pre-exponential
(\ref{norm}). The corrections are of the two different types

\beq\label{corr}
{\cal O} ({\partial \over \partial v_k} W) + "quantum \ \ corrections"
\eeq

\bigskip

First type should dissappear if one introduce a new
"spectral parameter"

\beq\label{mtilde}
W(M) = \tilde M^p.
\eeq

Thus, we are led to special time variables induced by a special transformation
of the spectral parameter $\mu$

\beq\label{miwatilde}
\tilde T_k = {1\over k} Tr \tilde M^{-k}
\eeq
The formula (\ref{corr}) also demonstrates that the $\{ v_k \}$ are not
the true time variables for a given arbitrary potential. Indeed, it
appears that the right variables are
the parameters  $\{t_k\}$  being
certain linear combinations of the coefficients  $\{v_k\}$  of the potential
\cite{DVV,Kri}
\beq\label{kri}
t_k = - {p\over k(p-k)}Res\ W^{1-k/p}(\mu )d\mu
\eeq

Using (\ref{kri}) one can get two important formulas:
\beq
\mu  =  {1\over p}\sum ^{p+1}_{-\infty } kt_k \tilde \mu ^{k-p},
\eeq
and
\beq\label{pref}
V(\mu ) - \mu V'(\mu ) = - \sum ^{p+1}_{-\infty } t_k \tilde \mu ^k.
\eeq
The second one implies the natural
interpretation of the exponential pre-factor in eq.(\ref{gkm})
as the standard essential
singularity factor in the Baker-Akhiezer function of $p$-time variables.

Now, the  direct calculation shows
that
\beq\label{10}
Z[V|T_k] = \tau _V[T_k] =
$$
$$
= \exp \left( - {1\over 2}\sum    A_{ij}(t)(\tilde T_i-t_i)(\tilde T_j-t_j)
\right)  \tau _p[\tilde T_k- t_k]\hbox{  ,}
\eeq
where
\beq\label{11}
A_{ij} = Res_\mu  W^{i/p} dW^{j/p}_+\hbox{ ,}
\eeq
and $f(\mu )_+$ denotes the positive part of the Laurent series $f(\mu ) =
\Sigma \ f_i\mu ^i$. It is also easy to demonstrate, that
\beq\label{S.1}
\tau _p[T] \equiv  \tau _{V_p}[T]
\eeq
- is the $\tau $-function of $p$-reduction.

Formula (\ref{10}) means that ``shifted" by flows along $p$-times
(\ref{kri})
$\tau $-function is
easily expressed through the $\tau $-function of  $p$-reduction, depending only
on the difference of the time-variables $\tilde T_k$ and $t_k$. The change of
the spectral parameter in (\ref{5})  $M \rightarrow  \tilde M = f(M) =
W^{1/p}(M)$  (and
corresponding
transformation of times  $T_k \rightarrow  \tilde T_k)$  is a natural step from
the point of view of equivalent hierarchies.

Indeed,
the relation between $\tau $-functions of the equivalent hierarchies can be
easily derived from an identical transformation:
\beq\label{S.15}
\tau (T) = {\Delta (\tilde \mu )\over \Delta (\mu )}
\prod_i
[f'(\mu _i)]^{1/2} \tilde \tau (\tilde T)
\eeq
where $\tilde \tau (\tilde T)$ as function of times $\tilde T$
has the determinant
form (\ref{det}) with the basic vectors
\beq\label{S.16}
\tilde \phi (\tilde \mu ) =
[f'(\mu (\tilde \mu ))]^{1/2}\phi _i(\mu (\tilde \mu ))
\eeq
By a direct calculation one can show that pre-factor in
eq.(\ref{S.15}) may be
represented in the form
\beq\label{S.17}
{\Delta (\tilde \mu )\over \Delta (\mu )}
\prod_i
[f'(\mu _i)]^{1/2} = \exp \left( - {1\over 2}
\sum _{i,j}A_{ij}\tilde T_i\tilde T_j\right)
\eeq
where
\beq\label{S.18}
A_{ij} = Res\ f^i(\lambda )d_\lambda f^j_+(\lambda ).
\eeq

{}From (\ref{S.15})
we see that
\beq\label{S.25}
\tau (T(\tilde T)) = \tilde \tau (\tilde T) \exp \left( - {1\over 2}
\sum _{i,j}A_{ij}\tilde T_i\tilde T_j\right)
\eeq
Let us introduce the
$\tau $-function $\hat \tau (\tilde T)$ of the $p$-reduced KP hierarchy
defined by:
\beq\label{S.26}
\tilde \tau (\tilde T) \equiv  {\hat \tau (\tilde T)\over \tau _0(t)}
\exp \left( \sum  _j jt_{-j}\tilde T_j\right)
$$
$$
\tau _0(t) = e^{-{1\over 2} \sum A_{ij}t_it_j}
\eeq
for which instead of
\beq\label{13}
\tau _V[T] = {\det \ \phi _i(\mu _j)\over \Delta (\mu )}
\eeq
we have
\beq\label{14}
{\tau _p[\tilde T-t]\over \tau _p[t]} = {\det
\hat \phi _i(\tilde \mu _j)\over \Delta (\tilde \mu )}
\eeq
with the corresponding points of the Grassmannian determined by the basic
vectors
\beq\label{15}
\phi _i(\mu ) = [W'(\mu )]^{1/2} \exp \left( V(\mu ) - \mu W(\mu )\right)
\int   x^{i-1}e^{-V(x)+xW(\mu )} dx
\eeq
and
\beq\label{16}
\hat \phi _i(\tilde \mu ) = [p\tilde \mu ^{p-1}]^{1/2} \exp \left(
-\sum ^{p+1}_{j=1}t_j\tilde \mu _j\right)  \int
x^{i-1}e^{-V(x)+x\tilde \mu ^p}dx
\eeq
respectively. Then it is easy to show that $\hat \tau _p(T)$ satisfies the
$L_{-1}$- constraint with {\it shifted} KP-times in the following way
\beq\label{17}
\sum ^{p-1}_{k=1}k(p-k)(\tilde T_k-t_k)(\tilde T_{p-k}-t_{p-k}) +
\sum ^\infty _{k=1}(p+k)(\tilde T_{p+k}-t_{p+k}){\partial \over \partial %
\tilde T_k} \log  \hat \tau _p[\tilde T-t] = 0
\eeq
where $t_i$ defined by (\ref{kri}) are {\it identically} equal to zero for
$i \geq
p+2.$

The formulas (\ref{10},\ref{17}) demonstrate at least two things. First, the
partition function
in the case of deformed monomial potential $(\equiv  polynomial$ of the same
degree) is expressed through the equivalent solution (in the sense
\cite{Shiota,Tak}) of the same $p$-reduced KP hierarchy, second -- not only
$t_{p+1}$ but all  $t_k$ with  $k \leq  p+1$  are not equal to zero in the
deformed situation. We will
call such theories as {\it topologically deformed $(p,1)$} models (in contrast
to {\it pure $(p,1)$} models given by monomial potentials  $V_p(X))$, the
deformation is ``topological" in the sense that it preserves all the features
of topological models. Moreover, this ``topological"
deformation preserves almost all features of $2d$ Landau-Ginzburg theories
and from the point of view of continuum theory they should be identified
with the twisted Landau-Ginzburg topological matter interacting with
topological gravity
\footnote{In spherical limit this conclusion was also made in
\cite{Losev}}.

(Of course, the basic vectors of the pure $(p,1)$ model corresponding to
monomial potential $V_p$ can be obtained by setting $t_1 = ..$. $= t_p = 0$,
$t_{p+1} = {p\over p+1} )$. As a solution to string equation this deformed case
differs only in analytic continuation along first  $p$  times.

These topologically deformed $(p,1)$ models as we already said preserve all
topological properties of $(p,1)$ models. Indeed, according to \cite{FKN1}
shifting
of first times  $t_1,...,t_{p+1}$ is certainly not enough to get higher
critical points. To do this one has to obtain  $t_{p+q} \neq  0$, but this
cannot be done using above formulas naively, because it is easily seen from
definition (\ref{kri}) of $p$-times, that  $t_k \equiv  0$  for  $k \geq
p+2$.

\section{General scheme and $pq$-duality}

The above scheme has a natural quasiclassical
interpretation. Indeed, the solution to $(p,1)$ theories given by the partition
function (\ref{gkm}) can be considered as a ``path integral" representation
of the
solution to Douglas equations \cite{Douglas}
\beq\label{20}
[\hat P,\hat Q] = 1
\eeq
where  $\hat P$  and  $\hat Q$  are certain differential operators (of order
$p$  and  $q)$ respectively and obviously  $p-th$ order of  $\hat P$  dictates
$p$-reduction, while  $q$  stands for  $q-th$ critical point. Quasiclassically,
(\ref{20}) turns into Poisson brackets relation  \cite{Kri,TakTak}
\beq\label{21}
\{P,Q\} = 1
\eeq
where  $P(x)$  and  $Q(x)$  are now certain (polynomial) functions. It is
easily seen that the above case corresponds to the first order polynomial
$Q(x) \equiv  x$  and the $p$-th order polynomial  $P(x)$  should be identified
with  $W(x)\equiv V'(x)$. Thus, the exponentials in
(\ref{gkm}), (\ref{15}) and (\ref{16}) acquire an obvious sense of action
functionals
\beq\label{22}
S_{p,1}(x,\mu ) = - V(x) + xW(\mu ) = - \int ^x_0dy\ W(y)Q'(y) + Q(x)W(\mu )
$$
$$
W(x) = V'(x) = x^p + \sum ^p_{k=1}v_kx^{k-1}
$$
$$
Q(x) = x
\footnote{For example: $W(z) = z^2 + t_1$, $Q(z)=z$, then for (\ref{21})
$$
\{ W,Q\} = {\partial W \over \partial t_1}{\partial Q \over \partial z} -
 {\partial Q \over \partial t_1}{\partial W \over \partial z} = 1
$$}
\eeq
and we claim that the generalization to arbitrary $(p,q)$ case must be
\beq\label{23}
S_{W,Q} =  - \int ^x_0dy\ W(y)Q'(y) + Q(x)W(\mu )
$$
$$
W(x) = V'(x) = x^p + \sum ^p_{k=1}v_kx^{k-1}
$$
$$
Q(x) = x^q + \sum ^q_{k=1}\bar v_kx^{k-1}
\eeq
Now the ``true" co-ordinate is $Q$, therefore the extreme condition of action
(\ref{23}) is still
\beq\label{24}
W(x) = W(\mu )
\eeq
having  $x = \mu $  as a solution, and for extreme value of the action one gets
\beq\label{25}
\left.S_{W,Q} \right |_{x=\mu } = \int ^\mu _0 dy\ W'(y)Q(y) =
$$
$$
= \sum ^{p+q}_{k=-\infty }t_k\tilde \mu ^k
\eeq
where  $\tilde \mu ^p = W(\mu )$  and
\beq\label{26}
t_k \equiv  t^{(W,Q)}_k = - {p\over k(p-k)}Res\ W^{1-k/p}dQ\hbox{  .}
\eeq
We should stress that the extreme value of the action (\ref{23}),
represented in the form (\ref{25}), determines the quasiclassical (or
dispersionless)
limit of the $p$-reduced KP hierarchy \cite{Kri,TakTak} with $p+q-1$
independent flows. We have seen that in the case of topologically deformed
$(p,1)$ models the quasiclassical hierarchy is exact in the strict sense:
topological solutions satisfy the full KP equations and the first basic vector
is just the Baker-Akhiezer function of our model (\ref{gkm}) restricted to the
small phase
space. Unfortunately, this is not the case for the general $(p,q)$ models: now
the quasiclassics is not exact and in order to find the basic vectors in the
explicit form one should solve the original problem and find the exact
solutions of the full KP hierarchy along first $p+q-1$ flows. Nevertheless,
 the presence of the ``quasiclassical component" in the whole
integrable structure of the given models can give, in
principle, some useful information, for example, we can make a conjecture that
the coefficients of the basic vectors are determined by the derivatives of the
corresponding {\it quasiclassical $\tau $}-function.

Returning to eq.(\ref{26}) we immediately see, that now only for  $k \geq
p+q+1$ \ \ $p$-times are identically zero, while
\beq\label{27}
t_{p+q} \equiv  t^{(W,Q)}_{p+q} = {p\over p+q}
\eeq
and we should get a correct critical point adjusting all  $\{t_k\}$  with  $k <
p+q$  to be zero. The exact formula for the Grassmannian basis vectors in
general case acquires the form

\beq\label{dual}
\phi _i(\mu ) = [W'(\mu )]^{1/2} \exp ( - \left.S_{W,Q}\right |
_{x=\mu })  \int   d{\cal M}_Q(x)f_i(x) \exp \ S_{W,Q}(x,\mu )
\eeq
where  $d{\cal M}_Q(x)$  is the integration measure. We are going to explain,
that the integration measure for generic theory determined by {\it two}
arbitrary
polynomials $W$  and  $Q$  has the form
\beq\label{29}
d{\cal M}_Q(z) = [Q'(z)]^{1/2}dz
\eeq
by checking the string equation. For the choice (\ref{29}) to insure the
correct
asymptotics of basis vectors  $\phi _i(\mu )$  we have to take  $f_i(x)$  being
functions (not necessarily polynomials) with the asymptotics
\beq\label{30}
f_i(x) \sim  x^{i-1}(1 + O(1/x))
\eeq

To satisfy the string
equation, one has to fulfill two requirements: the reduction condition
\beq\label{sch1}
W(\mu )\phi _i(\mu ) = \sum  _j C_{ij}\phi _j(\mu )
\eeq
and the Kac-Schwarz (\ref{ks}) operator action
\beq\label{sch2}
A^{(W,Q)}\phi _i(\mu ) = \sum    A_{ij}\phi _j(\mu )
\eeq
with
\beq\label{33}
A^{(W,Q)} \equiv  N^{(W,Q)}(\mu ){1\over W'(\mu )}
{\partial \over \partial \mu } [N^{(W,Q)}(\mu )]^{-1} =
$$
$$
= {1\over W'(\mu )} {\partial \over \partial \mu } - {1\over 2}
{W''(\mu )\over W'(\mu )^2} + Q(\mu )
$$
$$
N^{(W,Q)}(\mu ) \equiv  [W'(\mu )]^{1/2} \exp ( - \left.S_{W,Q}
\right |_{x=\mu })
\eeq
These two requirements are enough to prove string equation.
The structure of action immediately gives us that
\beq\label{34}
A^{(W,Q)}\phi _i(\mu ) = N^{(W,Q)}(\mu )\int   d{\cal M}_Q(z) Q(z)f_i(z)
\exp \ S_{W,Q}(z,\mu )
\eeq
and the condition (\ref{sch2}) can be reformulated as a $Q$-reduction property
of basis
$\{f_i(z)\}$
\beq\label{35}
Q(z)f_i(z) = \sum    A_{ij}f_i(z)
\eeq

Let us check now the reduction condition. Multiplying  $\phi _i(\mu )$  by
$W(\mu )$  and integrating by parts we obtain
\beq\label{35a}
W(\mu )\phi _i(\mu ) =
$$
$$
= N^{(W,Q)}(\mu )\int   d{\cal M}_Q(z) f_i(z)
{1\over Q'(z)} {\partial \over \partial z} [\exp \ Q(z)W(\mu )] \exp [-
\int ^z_0dy\ W(y)Q'(y)] =
$$
$$
= - N^{(W,Q)}(\mu )\int   d{\cal M}_Q(z) \exp [S_{W,Q}(z,\mu )] \left(
{1\over Q'(z)} {\partial \over \partial z} - {1\over 2} {Q''(z)\over Q'(z)^2} -
W(z) \right) f_i(z) \equiv
$$
$$
\equiv  - N^{(W,Q)}(\mu )\int   d{\cal M}_Q(z) \exp [S_{W,Q}(z,\mu )]
A^{(Q,W)}f_i(z)
\eeq
Therefore, in the ``dual" basis  $\{f_i(z)\}$  the condition (31) turns to be

\beq\label{35b}
A^{(Q,W)}f_i(z) = - \sum    C_{ij}f_j(z)
\eeq
with  $A^{(Q,W)} (\neq  A^{(W,Q)})$  being the ``dual" Kac-Schwarz operator
\beq\label{36}
A^{(Q,W)} = {1\over Q'(z)} {\partial \over \partial z} - {1\over 2}
{Q''(z)\over Q'(z)^2} - W(z)
\eeq

The representation (\ref{dual}), (\ref{29}) is an exact integral formula for
basis vectors
solving the $(p,q)$ string model. It has manifest property of $p-q$ duality (in
general $W-Q$), turning the $(p,q)$-string equation into the equivalent
$(q,p)$-string equation.

Now let us transform (\ref{dual}), (\ref{29}) into a little bit more explicit
$p-q$ form. As
before for $(p,1)$ models we have to make substitutions, leading to
equivalent KP solutions:
\beq\label{37}
\tilde \mu ^p = W(\mu )\hbox{, }    \tilde z^q = Q(z)
\eeq
Then we can rewrite (\ref{dual}) as
\beq\label{38}
\hat \phi _i(\tilde \mu ) = [p\tilde \mu ^{p-1}]^{1/2} \exp \left( -
\sum ^{p+q}_{k=1}t_k\tilde \mu ^k\right)  \int   d\tilde z
[q\tilde z^{q-1}]^{1/2} \hat f_i(\tilde z) \exp \ S_{W,Q}(\tilde z,\tilde \mu )
\eeq
where action is given now by
\beq\label{39}
S_{W,Q}(\tilde z,\tilde \mu ) =  - \left [ \int ^{\tilde z}_0d\tilde y
q\tilde y^{q-1}W(y(\tilde y)) \right ]_+  + \tilde z^q \tilde \mu ^p
$$
$$
= \sum_{k=1}^{p+q} \bar t_k \tilde z^k + \tilde z^q \tilde \mu ^p
\eeq
In new coordinates the reduction conditions are
\beq\label{40}
\tilde \mu ^p\hat \phi _i(\tilde \mu ) = \sum  _j
\tilde C_{ij}\hat \phi _j(\tilde \mu )
$$
$$
\tilde z^q\hat f_i(\tilde z) = \sum  _j \tilde A_{ij}\hat f_j(\tilde z)
\eeq
and for the Kac-Schwarz operators one gets conventional formulas
\cite{KSch,Sch,KMMMZ91b}
\beq\label{41}
\tilde A^{(p,q)} = {1\over p\tilde \mu ^{p-1}}
{\partial \over \partial \tilde \mu } - {p-1\over 2p} {1\over
{\tilde \mu ^p}}
 + {1\over p} \sum ^{p+q}_{k=1}kt_k\tilde \mu ^{k-p}
$$
$$
\tilde A^{(q,p)} = {1\over q\tilde z^{q-1}} {\partial \over \partial \tilde z}
- {q-1\over 2q} {1\over {\tilde z^
q}} + {1\over q} \sum ^{p+q}_{k=1}k\bar t_k\tilde z^{k-q}
\eeq
where for $(q,p)$ models we have introduced the ``dual" times:
\beq\label{S.6}
\bar t_k \equiv  t^{(Q,W)}_k = {q\over k(q-k)}Res\ Q^{1-k/q}dW
\eeq
in particularly, $\bar t_{p+q} = - {q\over p} t_{p+q} = - {q\over p+q}$ . Now
string equations give correspondingly
\beq\label{42}
\tilde A^{(p,q)}\hat \phi _i(\tilde \mu ) = \sum
\tilde A_{ij}\hat \phi _j(\tilde \mu )
$$
$$
\tilde A^{(q,p)}\hat f_i(\tilde z) = -\sum
\tilde C_{ij}\hat f_j(\tilde z)
\eeq

By these formulas we get a manifestation of $p-q$ duality if solutions to $2d$
gravity.

As a main result of formulas presented above, one may conclude that a
{\it generic} solution to $c \leq 1$ $2d$ gravity should be described by
{\it two} functions $W(x)$ and $Q(x)$.

\section{String field theory and $c \to 1$ limit}

Now we will make some remarks why this model could be considered as
an attempt of constructing a {\it string field theory}
or effective theory of string models. It is necessary to point out from the
beginning that by string field theory we would mean more than a conventional
definition as a field theory of functionals defined on string loops - it must
rather mean a sort of effective theory which gives all the solutions to
classical string equations of motion ($2d$ conformal field theories coupled to
$2d$
gravity) as its vacua and allows us to consider all of them on
equal footing (within the same Lagrangian framework) and maybe even describe
the
flows betveen different string vacua. Of course, it should reproduce
perturbation expansion around any of these vacua.
In this sense, the conventional string field theory was not true
effective model, because it contained a {\it fixed} set of variables which
correspond to a concrete vacuum (say, 26 free scalar fields). Therefore, it
doesn't have even {\it a priori} a possibility to make a flow to another
classical solution (maybe only except for some simple change of a background),
$i.e$. conventional string field theory might describe only some small
perturbation around given classical solution in terms of the coordinates
equivalent to the matter variables in the Polyakov path integral
(\ref{pol}).

Moreover, in conventional
approach even the perturbative expansion was ill-defined due
to the presence of tachyon in the spectrum or in other words due to the
instability of the classical solution.  The only chance to get a sensible
effective theory appears after we make sense to the non-perturbative
description, presented above, and existing up to now only in the case of
non-critical and moreover ``non-tachyonic" strings.
That means that what we know up to now is the only
case of ``highly-noncritical" string models where the total matter-gravity
central charge
\beq
c_{matter} + c_{gravity} = 26
\eeq
is ``dominated" by the contribution of 2d gravity. This is far from the case of
critical string $(c_{matter} = 26)$  and close to the case of pure
two-dimensional gravity. Unfortunately, up two now this is the only region
where it is possible to formulate string theory consistently and at least put
the question what is the internal principle which might allow one to choose
dynamically a string vacuum.

Now, the above scheme is a {\it string} field theory in a sense it
is highly related
to the underlying module space \cite{K1,K2}. That is actually the main
difference with ordinary {\it field} theory:
the
geometrical sense of construction above was motivated by
cell-decomposition of module space \cite{HZ,Pen,K1,K2}.

So, at the moment we have a sort of theory, describing various $(p,q)$
models coupled to $2d$ gravity beyond the perturbation theory, and
technically advocating to the equations in the space of coupling
constants, which determine the generation function for all the
correlators. Such function (being the logariphm of the $\tau $-function
cannot be defined {\it globally} in this space, and around each
''critical point" it has some sort of perturbative definition, which
should reproduce the original ''first-quantized" theory (\ref{pol}).
However, moving in this space from one solution to another naively one will
meet
the divergences of such expansion, then it is necessary to continue
analytically, and the piece we would get in such a way is exactly a
nonperturbative correction. For simplest $(p,1)$ topological theories,
this scheme can be described at the language of effective (''path") or
rather matrix integral with almost trivial integration measure, the
exact integral formulas for generic case involve more complicated
structures.

This scheme, in principle, should also be true for the ''barrier" $c=1$
solutions. However, a generic $c=1$ ''phase" is rather complicated
theory, and naive one-matrix formulation leads only to some highly
restricted $c=1$ cases.

These cases are mostly based on various manipulations with the Penner
model.
Indeed, the determinant form of Penner model \cite{HZ,Pen,DVaf}
partition function implies already
that for fixed values of times it is a Toda lattice tau-function in the sense
of and allows us to
apply to this case the the Toda theory
representation for Generalized Kontsevich models. Indeed, the solution to
the Penner model
\beq\label{penner}
{\cal Z} \sim  \det \ {\cal H}^{(\alpha )}_{ij}
\eeq
with
\beq
{\cal H}^{(\alpha )}_{ij} = \Gamma (\alpha +i+j-1)
\eeq
is nothing but a specific case of GKM.

Now one can easily introduce positive- and negative-times dependence in
(\ref{penner})
and then reconstruct  $\Phi ^{\{V\}}_k(z)$  from (\ref{phi-})
 Indeed,
\beq
h^{(\alpha )}_{ij} = {\cal H}^{(\alpha )}_{ij} = \Gamma (\alpha -1+i+j) =
$$
$$
= \int ^\infty _0{dy\over y} e^{-y} y^{\alpha -1+i+j} =
\oint \phi ^{(\alpha )}_i(z)z^j
\eeq
immediately gives
\beq\label{phic1}
\phi ^{(\alpha )}_i(z) = \int ^\infty _0{dy\over y} e^{zy-y} y^{\alpha -1+i}
\eeq
which is a sort of GKM-like representation. The difference with more common
situation for  $c<1$  is in the definition of the contour in
(\ref{phic1}) and also in
the fact that  $z$-dependence is trivial, because integral is easily
taken with the result
\beq
\phi ^{(\alpha )}_i(z) = {\Gamma (\alpha +i)\over (z-1)^{\alpha +i}} \equiv
\phi _{\alpha +i}(z)
\eeq
and
\beq
\left( {\partial \over \partial z}\right) ^j\phi ^{(\alpha )}_i(z) =
(-)^j\phi ^{(\alpha )}_{i+j}(z) = (-)^j
{\Gamma (\alpha +i+j)\over (z-1)^{\alpha +i+j}}
\eeq
Introducing negative times, one gets \cite{Mar92}
\medskip
\beq
\phi ^{(\alpha )}_i(z|T_{-p}) =
z^{-\alpha }\exp \left( -\sum _{p>0}T_{-p}z^{-p}\right) \sum  _k
P_k[T_{-p}]\phi ^{(\alpha )}_{i-k}(z)
\eeq
where  $P_k[T_p]$   are Schur polynomials  $(\exp \ \sum T_pz^p =
\sum \ z^nP_n[T_p]).$ or simply
\beq
Z_{c=1} \sim \int DY \exp {Tr ZY + \alpha Tr log Y +
\sum_{k>0} T_{-k} Tr Y^{-k}}
\eeq
with
\beq
T_{+k} = {1 \over k} Tr Z^k
\eeq
and amazingly this formula turns to be consistent with the calculations
of the tachyonic amplitudes \cite{DMP}.

Let us now make more comments on $c=1$ situation. From
basic point of view we need in generic situation to get the most general
(unreduced) KP or Toda-lattice tau-function satisfying some (unreduced) string
equation. In a sense this is not a limiting case for $c<1$ situation but rather
a sort of ``direct sum" for all $(p,q)$ models. This reflects that in conformal
theory coupled to $2d$ gravity there is, in a sense, less difference between
$c<1$ and $c=1$ situations than this coupling.

However, there are several particular cases when one can construct a sort of
direct $c \to 1$ limit and which should correspond to certain highly
``degenerate" $c=1$ theories. From the general point of view presented above
these are nothing but very specific cases of $(p,q)$ string equations, and
they could correspond only to a certain very reduced subsector of $c=1$ theory.

Indeed, it is easy to see, that for two special cases $p = \pm q$ the equations
(\ref{sch1}), (\ref{sch2})  can be simplified drastically, actually giving
rise to a single
equation instead of a system of them. Of course, these two cases don't
correspond to minimal series where one needs $(p,q)$ being coprime numbers.
However, we still can fulfill both reduction and Kac-Schwarz condition and
these solutions to our equations using naively the formula for the central
charge, one might identify with $c=1$ for $p=q$ and $c=25$ for $p=-q$.

Now, the simplest theories should be again with $q=1$. For
such case
``$c=1$" turns to be equivalent to a discrete matrix model \cite{KMMM92a}
while ``$c=25$" is exactly what one would expect from generalization of
the Penner approach \cite{Mar92,DMP}. Indeed, taking in general
{\it non-polynomial} functions, like
\beq
W(x) = x^{-\beta }
$$
$$
Q(x) = x^{\beta }
\eeq
the action would acquire a logariphmic term
\beq
S_{-\beta ,\beta } = - \beta logx + {x^{\beta } \over \mu^{\beta }}
\eeq
while equations (\ref{sch1}), (\ref{sch2}) give rise just to rational
solutions. It is very
easy to see that $\beta = 1$ immediately gives the Penner model in the
external field,
which rather corresponds to ``dual" to $c=1$ situation with matter central
charge being $c_{matter}=25$ with a highly non-unitary realization of
conformal matter
\footnote{This $c=1$ \ -- \ $c=25$ duality might be also connected with the
known fact that there exists a Legendre transform between the
Gross-Klebanov solution to c=1 matrix model and the Penner model}.

On the other hand, $p=q=1$ solution is nothing but a trivial theory,
which however becomes a nontrivial discrete matrix model for unfrozen
zero-time. Moreover, these particilar $p = \pm q$ solutions become nontrivial
only if one considers the Toda-lattice picture with negative times being
involved into dynamics of the effective theory. On the contrary, we know
that $c<1$ $(p,q)$-solutions in a sense trivially depends on negative times
with the last ones playing the role of symmetry of string equation
\cite{KMMM92a}. It means, that we don't yet understand
enough the role of zero and negative times in the Toda-lattice formulation.

\section{Conclusion}

Let us, finally make several concluding remarks. We tried to demonstrate
above that using additional structure of integrability arised in the
''discretized" formulation of $2d$ gravity one might go far from what is
known in string perturbation expansion. Namely, a generic $(p,q)$
solution to the $c < 1$ case can be described as a solution to the
integrable system and its explicit form can be studied using various
integral representations for these ''stringy" solutions. In the most
simplest case of topological theories, these integral formulas can be
combined into the {\it matrix} integral giving rise to a matrix
''zero-dimensional" field theory description.

Much more complicated is the case of $d \geq 1$ which at the moment
seems to be almost unclear from the point of view be discussed below.
One might only hope that there exists a possible generalization
concerning {\it non-Cartan} flows and formed by them an integrable
system. One might also hope that such integrable systems are related to
the matrix-like integrals with a nontrivial integration over ''angle
variables", $i.e.$ when in matrix integral (\ref{mamo}) the action is no
longer the trace of function of a matrix, but is rather complicated
objects. Such type of systems were discussed in the literature
\cite{ind.grav} and the technique was mostly based on the more
complicated operations with the Itzykson-Zuber integrals \cite{IZ}.
This might be connected to a sort of ''group-theoretical"
$\tau$-functions, to be discussed elsewhere \cite{taugr}.

\section{Acknowlegements}

I would like to thank J.Petersen and J.Sidenius who initiated the course
of lectures after which this paper has appeared. I am also indebted to
D.Boulatov, L.Chekhov, P.Di Vecchia, K.Ito, C.Kristjansen H.Nillsen and
especially
to J.Amb{\o}rn for valuable discussions and remarks. And I am certainly
grateful to A.Losev, S.Kharchev, A.Mironov, A.Morozov and A.Zabrodin for
collaboration and permanent discussions on presented above and related
problems. This work was financially supported by NORDITA.

\end{document}